An approach to first principles electronic structure calculation by symbolic-numeric computation

Akihito Kikuchi
CANON INC.
30-2, Shimomaruko 3-chome, Ohta-Ku, Tokyo 146-8501-Japan
E-mail:kikuchi.akihito@canon.co.jp

**ABSTRACT**
This article is an introduction to a new approach to first principles electronic structure calculation. The starting point is the Hartree-Fock-Roothaan equation, in which molecular integrals are approximated by polynomials by way of Taylor expansion with respect to atomic coordinates and other variables. It leads to a set of polynomial equations whose solutions are eigenstate, which is designated as algebraic molecular orbital equation. Symbolic computation, especially, Gröbner bases theory, enables us to rewrite the polynomial equations into more trimmed and tractable forms with identical roots, from which we can unravel the relationship between physical parameters (wave function, atomic coordinates, and others) and numerically evaluate them one by one in order. Furthermore, this method is a unified way to solve the electronic structure calculation, the optimization of physical parameters, and the inverse problem as a forward problem.

# INTRODUCTION

This article is intended for an introduction of a new approach to first principles electronic structure calculation by way of symbolic-numeric computation [1].

There is a wide variety of electronic structure calculation cooperating with symbolic computation. The latter is mainly purposed to play the auxiliary role (but not without importance) to the former. In the field of quantum physics [1-9], researchers sometimes have to handle with complicated mathematical expressions, whose derivation seems to be almost beyond human power. Thus one resorts to intensive use of computer, namely, symbolic computation [10-16]. The example concerning this, as is given in the reference 16, ranges various topics: atomic energy levels, molecular dynamics, molecular energy and spectra, collision and scattering, lattice spin models and so on. How to obtain molecular integrals analytically or how to manipulate complex formulas in many-body interaction, is one of such problems. In the former, when one uses special atomic basis for a specific purpose, to express the integrals by the combination of already-known analytic functions may sometimes be very difficult. In the latter, one must rearrange a number of creation and annihilation operators in a suitable order and calculate the analytical expectation value. In usual, a quantitative and massive computation is to follow from a symbolic one; for the convenience of the numerical computation, it is necessary to reduce a complicated analytic expression into a tractable and computable form. This is the main motive for the introduction of the symbolic computation as a forerunner of the numerical one and their collaboration has won considerable successes up to now. The present work should be classified as one of such trials. Meanwhile, the use of symbolic computation in the present work is not limited to indirect and auxiliary part to the numerical computation. The present work can be applicable to a direct and quantitative estimation of the electronic structure, skipping conventional computational methods.

The basic equation of the first principle electronic structure calculations is the Hartree-Fock or the Kohn-Sham equation, derived from the minimum condition of the energy functional in the electron-nuclei system [1-3], which is expressed as follows.

$$\left( -\frac{1}{2}\Delta + \sum_a \frac{Z_a}{|r - R_a|} + \int dr' \frac{\rho(r')}{|r - r'|} + V^{exc} \right) \phi_i(r) = E_i \phi_i(r). \quad (I.1)$$

The second term in the parse of the left side is the potential from nuclei with charge $Z_a$. The third term is the Coulomb potential generated by the charge distribution $\rho$. The forth term $V^{exc}$ is the quantum dynamical interaction operating in the many-electron system, called the "exchange and correlation". It is rewritten into the matrix eigenvalue problem, by adopting wavefunctions expanded by a certain localized basis set. This is Hartree-Fock-Roothaan equation in eq. (I.2) (hereafter abbreviated as HFR equation).

$$\mathbf{H}(\{R\},\{\Psi\},\{\zeta\},\{Q\}) \ \mathbf{\Psi}(\{R\},\{\zeta\},\{Q\}) = \mathbf{S}(\{R\},\{\zeta\},\{Q\}) \ \mathbf{\Psi}(\{R\},\{\zeta\},\{Q\}) \ \mathbf{E}(\{\zeta\},\{Q\}). \quad (I.2)$$

**H** is the Hamiltonian matrix, **S** is the overlap one, **Ψ** is the wavefunction (the coefficients of the linear combination), and **E** is the eigenvalue. The variables {R} are the positions of the nuclei, {ζ} are the orbital exponents that describe the special expansion of the localized base function, and {Q} are the quantum numbers. The localized atomic bases are parameterized by them. The HFR equation can be expressed by multi-valuable analytic functions, whose variables are {R}, {ζ} and {Q}. It contains transcendental functions of several kinds. This is because we conventionally adopt analytical base, such as STO (Slater type orbital) or GTO (Gaussian type orbital), to construct one-electron and two-electron molecular integrals, whose concrete expression can be derived from symbolic manipulation [4-9] by means of computer algebra systems [10, 11].

The use of analytic basis in the HFR equation is effective in the achievement in the precision of the numerical computation, but causes some difficulty in the mathematical operations to the equation itself, because the analytic expression is, in general, very complicated. We, however, can rewrite and approximate the HFR equation by polynomials in order to obtain much simpler expressions. The concept of the polynomial approximation to the HFR equation is nourished by Yasui [6-9]. The equation becomes the set of algebraic polynomial equations expressed by atomic coordinates, orbital exponents, and quantum numbers. Based on this, we will able to unravel the relationship among parameters and clarify their dependence. This idea should be outlined here. We express a molecular orbital by the linear combination of atomic orbitals (LCAO),

$$\phi_k = \sum_\alpha^{atom} \sum_i^{base} C_{ai}^k \chi(R_\alpha, \{n_i, l_i, m_i\}, \zeta_i, r, \theta, \phi). \qquad (I.3)$$

The variables $R_\alpha, \{n_i, l_i, m_i\}, \zeta_i$ are the atomic position, quantum numbers, and the orbital exponents respectively. The variables $r, \theta, \phi$ are of the atom centered coordinates. The key components to the molecular orbital calculations are molecular integrals, which are the matrix elements of the each part of the Hamiltonian operator obtained by the use of LCAO, such as, the kinetic energy, the nuclear and electronic potentials, and the overlapping integrals. The approximation to molecular integrals is obtained by Taylor expansion;

$$f(x) = \sum_{p=0}^\infty \frac{1}{p!} \sum_k^p \binom{p}{k} (x-x_0)^{p-k} f^{(p)}(x_0) x^p \cong \sum_{i=0}^N A_i(x_0) x^i. \qquad (I.4)$$

For example, the two-centered overlapping integral is defined as,

$$S_{AB}^{ab}$$

$$\equiv \int \chi(R_A, \{n_a, l_a, m_a\}, \zeta_a, r_A, \theta_A, \phi_A) \, \chi(R_B, \{n_b, l_b, m_b\}, \zeta_b, r_B, \theta_B, \phi_B) dr^3. \quad (I.5)$$

The integration generates an analytic function of two orbital exponents and inter-atomic distance R. The polynomial approximation is given as,

$$S_{AB}^{ab}(\zeta_a, \zeta_b, R) \cong \sum_{p_a, p_b, p_R} A(\{n_a, l_a, m_a\}_A, \{n_b, l_b, m_b\}_B)_{p_a, p_b, p_c} \zeta_a^{p_a} \zeta_b^{p_b} R^{p_R}. \quad (I.6)$$

The other molecular integrals can be expressed in the similar way. Once the molecular integrals are approximated as polynomials, the HFR equation and the energy functional take polynomial expressions. The orbital exponents and atomic coordinates can equivalently be regarded as parameters in the calculus of variations, as well as LCAO coefficients. It will be adequate to call this multi-variable polynomial expression "algebraic molecular orbital equation," or "algebraic molecular orbital theory." It is not necessary to regard those equations as pure numerical eigenvalue problems. Those equations are a set of polynomials, to which both symbolic manipulations and numerical solving are applicable.

One should note some inconveniency in the conventional methods, which may be surmounted by "algebraic molecular orbital equation." The standard electronic structure calculation is a "forward problem". We suppose the material structure, execute the electronic structure calculations, and optimize the structure so that the energy functional will be minimized. The foundation for this treatment is so-called "the adiabatic approximation", which enables us to separate the dynamics of the nuclei and wavefunctions into two independent models, ruled by the classical and quantum dynamics respectively. The conventional method iterates two alternative computational phases, one of which are the optimization for the wavefunctions and the other for the positions of the nuclei. It is believed that this way is numerically stable. But, in view of effectiveness, this may be a lengthy and roundabout one. This also results in some inefficiency. Owing to the separation of the degrees of freedom of wavefunctions and nuclei, it is difficult for the conventional method to coop with cases where the dynamics of nuclei and wavefunctions are strongly coupled with each other. Meanwhile, the "inverse problem" will be to search the material structure which shows the desirable electronic properties. To do this, the conventional method must go with trial and error. At first we suppose the material structure to evaluate the electronic properties, and, by adjusting the structure, we search the direction in which the desired properties will be obtained. We have to solve forward problems repeatedly to obtain the solution of the inverse problems. The reason to this is as follows. In the conventional methods, the computation has the fixed order of numerical procedures, consisted from the eigenvalue solution, the self-consistent-field calculation and the relaxation of atomic structure, which is implemented as nested loops of independent phases of the optimizations. The unknown parameters are computed from inner loops to outer ones in order. The conventional method is obliged to determine unknown variables in a fixed order in any cases. As we will see later, the concept "algebraic molecular orbital equation" suggests a solution strategy to this circumstance.

**METHOD**

With the view of these circumstances, we propose the following method, named "Symbolic-numeric ab-initio molecular dynamics and molecular orbital method" [1].

It is summarized as follows. "At first, HFR equation is approximated as a set of multi-variable polynomial equations (Algebraic Molecular Orbital Equation), and by the symbolic computation, it is rewritten into a certain form more convenient for the numerical treatment. The eigenstates are evaluated by the root finding by means of symbolic-numeric procedure."

The question in the present work is how to obtain the numerical solutions of the equations after the polynomial approximation and derive the significant information. For the purpose of rewriting and solving a set of polynomial equations, the several types of hybrid techniques, so-called "symbolic-numeric solving", are proposed. In them, the symbolic manipulation is applied as a preconditioning toward the set of equations to be solved. The equations are transformed into the other ones which have the same roots, to which the numerical computation will be easy and stable. From the form of the transformed equations, the character of the solution, such as, the existence and the geometrical structure, can be determined. Then the solving process is passed over to the numerical one. For the mathematical background, see ref. [12-15]. A review of application of symbolic computations in the field of the computational chemistry is given in ref .16.

In the present work, we make use of the symbolic-numeric solving and rewriting HFR equation, approximated as a set of polynomial equations. As a strategy, the algorithm of the "decomposition of polynomial equations into triangular sets" is applied [12, 13]. In this algorithm, the following transformations are applied.

The starting equations $f_1, \ldots, f_n$

$$f_1(x_1, x_2, \ldots, x_n) = 0 \tag{M.1}$$
$$f_2(x_1, x_2, \ldots, x_n) = 0$$
$$\vdots$$
$$f_n(x_1, x_2, \ldots, x_n) = 0$$

→ Gröbner bases with lexicographic order of $f_1, \ldots, f_n, \{g_i\}$

$$g_1(x_1) = 0 \tag{M.2}$$
$$\vdots$$
$$g_{2\_1}(x_1, x_2) = 0$$
$$\vdots$$
$$g_{2\_m(2)}(x_1, x_2) = 0$$
$$g_{3\_1}(x_1, x_2, x_3) = 0$$
$$\vdots$$
$$g_{n\_1}(x_1, \ldots, x_n) = 0$$
$$\vdots$$
$$g_{n\_m(n)}(x_1, \ldots, x_n) = 0$$

→   Triangular sets of polynomials $\{t_i\}$, each of which is given as this.

$$t_1(x_1) = 0 \tag{M.3}$$
$$t_2(x_1, x_2) = 0$$
$$\vdots$$
$$t_n(x_1, x_2, \ldots, x_n) = 0$$

The algorithm in ref .12 and 13, at first, generates the Gröbner bases $\{g_i\}$ with the lexicographic monomial order from the starting set of equations $f_1, \ldots, f_n$. The generated Gröbner bases have roots identical to those of $f_1, \ldots, f_n$, and take forms which guarantee an easier numerical solving. The Gröbner bases are a set of polynomials in which the number of unknowns of each entry increases in order, from polynomials with fewer valuables to ones with more. However, the total number of polynomials in the Gröbner bases may grow more than that of the starting polynomials. Though we can search the root at this stage, we furthermore apply the decomposition to the Gröbner bases and obtain several "triangular" systems of equations $\{t_1, \ldots, t_n\}$, the first entry of which has one unknown $x_1$, the second has two unknowns $x_1, x_2$, the third three unknowns, etc, until, the last n-th has n-unknowns $x_1, x_2, \ldots, x_n$ by turns.   In order to obtain all roots of the starting equations $f_1, \ldots, f_n$, we may need to construct several sets of triangular set of equations, all of which can be generated by the algorithm in ref .12 or 13. Single triangular system has fewer numbers of entries than that in the Gröbner bases before the transformation. This makes the numerical procedure easier. Once we can obtain the triangular systems of the equations, we can evaluate each unknown one by one. In the numerical solution, only a Quasi-Newton-like method, or its kindred for one variable, is necessary.

We can execute another type of the symbolic numerical solving. The foundation to this is the theorem of Stickelberger. As above, we regard the HFR equations as the set of polynomial equations expressed by unknowns $X_1, \ldots, X_m$. The set of polynomial equations constructs a zero-dimensional ideal I in the polynomial ring $R=k[X_1, \ldots, X_m]$, whose zeros corresponds to a residue ring $A=R/I$. Here k is the coefficient field, which, in our cases, is the rational number field or the real number field. The ring A is a finite dimensional vector space over k, whose bases are expressed by monomials of $X_1, X_2, \ldots, X_m$. Thus, the multiplication by $X_1, X_2, \ldots, X_m$ on each base results in the linear combination of the bases in A. The product operations by $X_1, X_2, \ldots, X_m$ are expressed as linear transformation matrices $m_h$ ( $h = X_1, X_2, \ldots, X_m$). The bases in the residue ring A=R/I and the transformation matrices are obtained by means of Gröbner bases technique. The theorem of Stickelberger asserts that there is a one-to-one correspondence between an eigenvector $v_\xi$ of the matrix $m_h$ and the zero point $\xi = (\xi_1, \ldots, \xi_m)$ of the ideal I. The correspondence is given by

$$m_{X_i} \cdot v_\xi = \xi_i \cdot v_\xi. \qquad (M.4)$$

The eigenvector $v_\xi$ is common with all of $m_{X_i}$. (For details, see ref.14, p101-130, "From Enumerative Geometry to Solving Systems of Polynomial Equations", by Frank Sottile.) The numerical calculations for the zeros $\xi = (\xi_1, \ldots, \xi_m)$ are executed as follows. We choose one $X_i$ and prepare $m_{X_i}$. The secular equation gives us the eigenvector $v_\xi$. If we multiply $m_{X_j}$ (j ≠ i) with $v_\xi$, we can evaluate $\xi_j$ (j ≠ i). Thus all values of $\xi = (\xi_1, \ldots, \xi_m)$ can be obtained.

The energy functional, the normalization conditions for wavefunctions, and the HFR equation are polynomials with respect to LCAO (Linear Combination of Atomic Orbitals) coefficients and eigenvalues. Those equations are constructed from molecular integrals, which are in general expressed as analytic functions of the included parameters, accordingly not being polynomials. Thus, the molecular integrals are replaced by approximations of polynomials of the included parameters. By means of this, we can construct a set of polynomial equations including not only wavefunctions and eigenvalues but also the parameters for molecular integrals, i.e. the molecular orbital algebraic equation. If extra constraint conditions should be cast upon the HFR equation, we can prepare the polynomial equations for the constraints and add them into the set of polynomial equations. If the numerical coefficients are rationalized, we can avoid the lowering of the precision through the symbolic manipulation by means of arbitrary precision calculations of rational numbers.

The one of the merits in this treatment is as follows. In the conventional method, the input data is the atomic structure and the output if the electronic structure. By contrast, in the present method, the possible input data are not limited to the atomic structures. We can select arbitrary parameters in the HFR equation and set them as the input. If the problem to be solved is properly established, we can compute other unknown variables properly. As to the properness of the problem, i.e., the existence of the roots of the set of polynomial equations, it can be judged from the ideal theory in mathematics on the condition whether its Gröbner bases have zero points set or not.

### COMPUTATIONAL FLOW

The task flow is listed as follows.
1. Compute the analytic formula of the energy functional and the constraint conditions

whose variables are eigenvalues, LCAO coefficients in wavefunctions, atomic coordinates, and orbital exponents in molecular integrals and so on. Those analytic expressions are polynomials with respect to LCAO coefficients and eigenvalues while the expressions are not polynomials with respect to other parameters, such as atomic coordinates and orbital exponents. Molecular integrals are rewritten by approximating polynomials with respect to the included parameters. For example, by choosing a certain point in the range of a parameter and applying the Taylor expansion around it, we can obtain the polynomial approximation. By rationalizing the numerical coefficients, we can prevent the lowering of the precision through the afterward symbolic manipulation.

2. Prepare the set of the equations to be solved, by way of the symbolic derivative, which is the minimization of the energy functional approximated by the polynomial.

3. By means of the symbolic manipulation, the above equations are transformed into other ones having the same roots. The initial equations are, at first, transformed into the Gröbner bases, by which we can check the existence of the solution. If the solutions are the set of isolated points, we can decompose the Gröbner bases into the triangular expression. There is an alternative way; by means of Stickelberger's theorem, the search for the solution is replaced by an eigenvalue problem.

4. By numerically solving the equations, the roots are computed and afford us the electronic structure and the useful information.

## RESULTS

Several examples are demonstrated in this section. The units are given in atomic units. As an object, we choose the hydrogen molecule. Though we only show the examples of $H_2$ here, which is the simplest molecule, the applications of the present method are not limited to two-electron or two-atomic systems. The reason why we choose $H_2$ is as follows. Though this system is simple, it includes all kinds of quantum interactions operating in realistic materials and can be assumed as a miniature of general many-electron and polyatomic systems.

### To construct the algebraic molecular orbital equation

At first, we will show how to construct the algebraic molecular orbital equation and the possibility of the SCF (Self-Consistent Field) calculations. The molecular integrals needed here are generated by STO (Slater Type Orbital) base. The energy functional is the analytic equation of the one centered molecular integrals at hydrogen A and B, the two-centered molecular integrals, and the LCAO coefficients of the wavefunctions. As the expression of the inter-atomic distance R, the molecular integrals contain transcendental functions. One of the two-centered molecular integrals is shown in (R.1).

$$[1s(A)1s(A)|1s(B)1s(B)] \tag{R.1}$$

$$= \iint \mathrm{drdr'} \frac{\phi^{1s}(r-R_A;za)\phi^{1s}(r-R_A;zb)\phi^{1s}(r'-R_B;zc)\phi^{1s}(r'-R_B;zd)}{|r-r'|} =$$

$$= \frac{64\, za^{3/2}zb^{3/2}zc^{3/2}zd^{3/2}}{R(za+zb)^3(zc+zd)^3}$$

$$- \frac{32\, za^{3/2}zb^{3/2}(za+zb)zc^{3/2}zd^{3/2}}{E^{R(zc+zd)}(za+zb-zc-zd)^2(zc+zd)^2(za+zb+zc+zd)^2}$$

$$- \frac{32\, za^{3/2}zb^{3/2}zc^{3/2}zd^{3/2}(zc+zd)}{E^{R(za+zb)}(za+zb)^2(za+zb-zc-zd)^2(za+zb+zc+zd)^2}$$

$$- \frac{64\, za^{3/2}zb^{3/2}zc^{3/2}zd^{3/2}(zc+zd)(3za^2+6za\,zb+3zc^2-2zc\,zd-zd^2)}{E^{R(za+zb)}R(za+zb)^3(za+zb-zc-zd)^3(za+zb+zc+zd)^3}$$

$$+ \frac{64\, za^{3/2}zb^{3/2}(za+zc)zc^{3/2}zd^{3/2}(-za^2-2za\,zb-3zc^2+6zc\,zd+3zd^2)}{E^{R(zc+zd)}R(za+zb-zc-zd)^3(zc+zd)^3(za+zb+zc+zd)^3}$$

It is the two electron repulsion between 1s orbitals and classified as "Coulombic type", denoted as $[1s(A)1s(A)|1s(B)1s(B)]$. Here the notation "1s (A)" and "1s (B)" mean the atomic orbitals centered on atoms A and B. The Slater orbital takes a form $\phi^{1s}(r;z) = \frac{z^{3/2}}{\pi^{1/2}}e^{-zr}$. There are other repulsion integrals, classified as the "exchange type" $[1s(A)1s(B)|1s(A)1s(B)]$ and the "hybrid type" $[1s(A)1s(A)|1s(A)1s(B)]$. In general, the molecular integrals have more complicated expressions than (R.1). If STO is used, these integrals take more lengthy expressions, including transcendental functions, such as exponentials or exponential integrations, and infinite series summations [5]. In spite of this complicacy, we can treat them easily after the symbolic processing, by approximating them as finite degree polynomials by means of the Taylor expansion. It is noted here that the STO base can describe the physical property of the localized atomic wavefunction more precisely than by GTO (Gaussian Type Orbital) base, both in the neighborhood of the nucleus and in the remote region from it. Thus the STO base becomes more advantageous for the purpose of expressing the molecular equations as the polynomials of the atomic coordinates. This is the reason why STO is adopted here. However, the following recipes are also applicable to the GTO calculations, and possibly, to semi-empirical calculations, such as AM1 (Austine Model 1) [17] and PM3 (Parameterized Model number 3) [18], or tight-binding model, where the matrix elements are given as analytic formulas.

As an example of a forward problem in the first principles molecular dynamics, the optimization of the structure (the distance between two hydrogen atoms) and the UHF electronic structure calculation are simultaneously executed.

We execute the UHF (Unrestricted Hartree-Fock) calculations in which the trial wavefunctions for up- and down-spins are defined as in (R.2) and (R.3). The corresponding eigenvalues are denoted as ev and ew. It is noted here: these trial functions with the bases of the orbital exponent 1 are too primitive to assure good agreements with experiments: for accuracy, we must optimize the orbital exponent to a suitable value. It is only to reduce the computational cost in the symbolic computation why we adopt such primitive trial functions. Here, the positions of hydrogen A and B are denoted as $R_A$ and $R_B$. The inter-atomic distance is $r = R_A - R_B$. We use the notation, such as $x_A = |x - R_A|$ and $x_B = |x - R_B|$.

$$\phi_{up}(x) = (a \exp(-x_A) + b \exp(-x_B))/\sqrt{\pi} \tag{R.2}$$

$$\phi_{down}(x) = (c \exp(-x_A) + d \exp(-x_B))/\sqrt{\pi} \tag{R.3}$$

The energy functional is transformed into the polynomial form by way of the fourth order Taylor expansion of the inter-atomic distance r, centered at the position of $R_0$=7/5 atomic unit. The functional is generated in a standard way of molecular orbital theory, which is the total energy of the electron-nuclei system (given in the atomic units) with the constraint condition of the ortho-normality of the wavefunctions. The Lagrange multipliers are eigenvalues. The coefficients of real numbers are truncated to third decimal places and approximated as rational numbers, as is shown in (R.4). (a,b) are the LCAO coefficients for the up-spin electron, (c,d) are those for the down-spin electron, ev is the eigenvalues of the up-spin electron, ew is that of the down-spin electron, r is the inter-atomic distance. It is rather a rough approximation to use numerical coefficients truncated to third decimal places, which causes the computational error (the order of a few percents) by the present method, compared to the conventional way with the double precision calculation. This approximation is only intended to reduce the computational cost in the symbolic processing. In this sense, examples in this section are mock-ups for realistic calculations, the aim of which is to illustrate the application of the present method, aside from the accuracy.

(R.4)

$$\Omega[\{\phi_i(\xi); \text{the occupied orbital } i, \xi \equiv (r, \sigma_{spin})\}]$$

$$= \sum_i \int d\xi\, \phi_i(\xi) \left(-\frac{1}{2}\nabla^2 + \sum_a \frac{Z_a}{|r - R_a|}\right) \phi_i(\xi)$$

$$+ \frac{1}{2}\sum_{i,j} \iint d\xi d\xi' \frac{\phi_i(\xi)\phi_i(\xi)\phi_j(\xi')\phi_j(\xi')}{|r - r'|}$$

$$- \frac{1}{2}\sum_{i,j} \iint d\xi d\xi' \frac{\phi_i(\xi)\phi_j(\xi)\phi_j(\xi')\phi_i(\xi')}{|r - r'|}$$

$$+ \frac{1}{2}\sum_{a,b(a\neq b)} \frac{Z_a Z_b}{|R_a - R_b|}$$

$$- \sum_{i,j} \lambda_{ij} \left(\int d\xi \phi_i(\xi)\phi_j(\xi) - \delta_{ij}\right)$$

⇓

Ω＝(3571 - 1580*a^2 - 3075*a*b - 1580*b^2 - 1580*c^2 + 625*a^2*c^2 + 1243*a*b*c^2 + 620*b^2*c^2 - 3075*c*d + 1243*a^2*c*d + 2506*a*b*c*d + 1243*b^2*c*d - 1580*d^2 + 620*a^2*d^2 + 1243*a*b*d^2 + 625*b^2*d^2 + 1000*ev - 1000*a^2*ev - 1986*a*b*ev - 1000*b^2*ev + 1000*ew - 1000*c^2*ew - 1986*c*d*ew - 1000*d^2*ew - 5102*r + 332*a^2*r + 284*a*b*r + 332*b^2*r + 332*c^2*r + 43*a*b*c^2*r + 20*b^2*c^2*r + 284*c*d*r + 43*a^2*c*d*r + 80*a*b*c*d*r + 43*b^2*c*d*r + 332*d^2*r + 20*a^2*d^2*r + 43*a*b*d^2*r - 63*a*b*ev*r - 63*c*d*ew*r + 3644*r^2 + 75*a^2*r^2 + 724*a*b*r^2 + 75*b^2*r^2 + 75*c^2*r^2 - 401*a*b*c^2*r^2 - 124*b^2*c^2*r^2 + 724*c*d*r^2 - 401*a^2*c*d*r^2 - 1372*a*b*c*d*r^2 - 401*b^2*c*d*r^2 + 75*d^2*r^2 - 124*a^2*d^2*r^2 - 401*a*b*d^2*r^2 + 458*a*b*ev*r^2 + 458*c*d*ew*r^2 - 1301*r^3 - 69*a^2*r^3 - 303*a*b*r^3 - 69*b^2*r^3 - 69*c^2*r^3 + 146*a*b*c^2*r^3 + 42*b^2*c^2*r^3 - 303*c*d*r^3 + 146*a^2*c*d*r^3 + 618*a*b*c*d*r^3 + 146*b^2*c*d*r^3 - 69*d^2*r^3 + 42*a^2*d^2*r^3 + 146*a*b*d^2*r^3 - 139*a*b*ev*r^3 - 139*c*d*ew*r^3 + 185*r^4 + 12*a^2*r^4 + 39*a*b*r^4 + 12*b^2*r^4 + 12*c^2*r^4 - 17*a*b*c^2*r^4 - 4*b^2*c^2*r^4 + 39*c*d*r^4 - 17*a^2*c*d*r^4 - 86*a*b*c*d*r^4 - 17*b^2*c*d*r^4 + 12*d^2*r^4 - 4*a^2*d^2*r^4 - 17*a*b*d^2*r^4 + 13*a*b*ev*r^4 + 13*c*d*ew*r^4)/1000

The precision of this approximation should be checked at first. The values of the original energy functional and the polynomial approximation at a = b = c = d = 1 and ev = ew = 0 are plotted in Fig.1 as the function of inter-atomic distance. It shows sufficient agreement in the range of r=1〜2 atomic unit. However, if the r goes out of this region, the polynomial approximation is not appropriate and we must try another center of the expansion $R_0$. By increasing the maximum degree of the Taylor expansion, we can enlarge the range of r where the polynomial approximation is valid.

We make use of the symmetry in $H_2$, and we express the wavefunctions as the linear combination of the symmetric and asymmetric ones, as in (R.5) and (R.6).

$$\phi_{up}(x) = t(\exp(-x_A) + \exp(-x_B))/\sqrt{\pi} + s(\exp(-x_A) - \exp(-x_B))/\sqrt{\pi} \quad \text{(R.5)}$$

$$\phi_{down}(x) = u(\exp(-x_A) + \exp(-x_B))/\sqrt{\pi} + v(\exp(-x_A) - \exp(-x_B))/\sqrt{\pi} \quad \text{(R.6)}$$

This is the transformation by (R.7).

$$a = t + s, b = t - s, c = u + v, d = u - v \quad \text{(R.7)}$$

Then the HFR equation is given by the set of equations in (R.8), where (t,s) is the

LCAO coefficient for up-spin, (u,v) is that for down spin, ev is the eigenvalue for up spin, ew is that for down spin, and r is the inter-atomic distance.

① $\dfrac{\partial \Omega}{\partial a} + \dfrac{\partial \Omega}{\partial b} = 0 \rightarrow$ (R.8)

32*s*u*v*r^4-336*s*u*v*r^3+992*s*u*v*r^2-160*s*u*v*r+40*s*u*v-324*t*u^2*r^4+2572*t*u^2*r^3-6448*t*u^2*r^2+584*t*u^2*r+19936*t*u^2+156*t*v^2*r^4-1068*t*v^2*r^3+2248*t*v^2*r^2-80*t*v^2*r-32*t*v^2+26*t*ev*r^4-278*t*ev*r^3+916*t*ev*r^2-126*t*ev*r-7972*t*ev+126*t*r^4-882*t*r^3+1748*t*r^2+1896*t*r-12470*t=0

② $\dfrac{\partial \Omega}{\partial a} - \dfrac{\partial \Omega}{\partial b} = 0$

→156*s*u^2*r^4-1068*s*u^2*r^3+2248*s*u^2*r^2-80*s*u^2*r-32*s*u^2-52*s*v^2*r^4+236*s*v^2*r^3-32*s*v^2*r^2-104*s*v^2*r+48*s*v^2-26*s*ev*r^4+278*s*ev*r^3-916*s*ev*r^2+126*s*ev*r-28*s*ev-30*s*r^4+330*s*r^3-1148*s*r^2+760*s*r-170*s+32*t*u*v*r^4-336*t*u*v*r^3+992*t*u*v*r^2-160*t*u*v*r+40*t*u*v=0

③ $\dfrac{\partial \Omega}{\partial c} + \dfrac{\partial \Omega}{\partial d} = 0$

→156*s^2*u*r^4-1068*s^2*u*r^3+2248*s^2*u*r^2-80*s^2*u*r-32*s^2*u+32**t*v*r^4-336*s*t*v*r^3+992*s*t*v*r^2-160*s*t*v*r+40*s*t*v-324*t^2*u*r^4+2572*t^2*u*r^3-6448*t^2*u*r^2+584*t^2*u*r+19936*t^2*u+26*u*ew*r^4-278*u*ew*r^3+916*u*ew*r^2-126*u*ew*r-7972*u*ew+126*u*r^4-882*u*r^3+1748*u*r^2+1896*u*r-12470*u=0

④ $\dfrac{\partial \Omega}{\partial c} - \dfrac{\partial \Omega}{\partial d} = 0$

→-52*s^2*v*r^4+236*s^2*v*r^3-32*s^2*v*r^2-104*s^2*v*r+48*s^2*v+32*s*t*u*r^4-336*s*t*u*r^3+992*s*t*u*r^2-160*s*t*u*r+40*s*t*u+156*t^2*v*r^4-1068*t^2*v*r^3+2248*t^2*v*r^2-80*t^2*v*r-32*t^2*v-26*v*ew*r^4+278*v*ew*r^3-916*v*ew*r^2+126*v*ew*r-28*v*ew-30*v*r^4+330*v*r^3-1148*v*r^2+760*v*r-170*v=0

⑤ $\dfrac{\partial \Omega}{\partial (\mathrm{ev})} = \langle \phi_{up} | \phi_{up} \rangle - 1 = 0$

→-13*s^2*r^4+139*s^2*r^3-458*s^2*r^2+63*s^2*r-14*s^2+13*t^2*r^4-139*t^2*r^3+458*t^2*r^2-63*t^2*r-3986*t^2+1000=0

⑥ $\dfrac{\partial \Omega}{\partial (\mathrm{ew})} = \langle \phi_{down} | \phi_{down} \rangle - 1 = 0$

→13*u^2*r^4-139*u^2*r^3+458*u^2*r^2-63*u^2*r-3986*u^2-13*v^2*r^4+139*v^2*r^3-458*v^2*r^2+63*v^2*r-14*v^2+1000=0

⑦ $\dfrac{\partial \Omega}{\partial r} = 0$

→312*s^2*u^2*r^3-1602*s^2*u^2*r^2+2248*s^2*u^2*r-40*s^2*u^2-104*s^2*v^2*r^3+354*s^2*v^2*r^2-32*s^2*v^2*r-52*s^2*v^2-52*s^2*ev*r^3+417*s^2*ev*r^2-916*s^2*ev*r+63*s^2*ev-60*s^2*r^3+495*s^2*r^2-1148*s^2*r+380*s^2+128*s*t*u*v*r^3-1008*s*t*u*v*r^2+1984*s*t*u*v*r-160*s*t*u*v-648*t^2*u^2*r^3+3858*t^2*u^2*r^2-6448*t^2*u^2*r+292*t^2*u^2+312*t^2*v^2*r^3-1602*t^2*v^2*r^2+2248*t^2*v^2*r-40*t^2*v^2+52*t^2*ev*r^3-417*t^2*ev*r^2+916*t^2*ev*r-63*t^2*ev+252*t^2*r^3-1323*t^2*r^2+1748*t^2*r+948*t^2+52*u^2*ew*r^3-417*u^2*ew*r^2+916*u^2*ew*r-63*u^2*ew+252*u^2*r^3-1323*u^2*r^2+1748*u^2*r+948*u^2-52*v^2*ew*r^3+417*v^2*ew*r^2-916*v^2*ew*r+63*v^2*ew-60*v^2*r^3+495*v^2*r^2-1148*v^2*r+380*v^2+740*r^3-3903*r^2+7288*r-5102=0

**To solve the algebraic molecular orbital equation by symbolic-numeric computation**

At first, we assume r=7/5 to see the possibility of actual first principles calculations. The entry as is shown by $\dfrac{\partial \Omega}{\partial r} = 0$ in (R.8) is replaced by (R.9).

$$5r - 7 = 0 \tag{R.9}$$

The lexicographic order Gröbner bases are shown in (R.10) with the monomial ordering of s<t<u<v<ev<ew<r. (Hereafter this monomial ordering is kept in the following computations.) In the entries of the Gröbner bases, those variables show themselves in the reverse order of r, ew, ev, v, u, t, s, by turns. The relationships among those variables, which are ambiguous in the expression of HFR equation, may be extracted there.

J[1]=r-1.4 (R.10)
J[2]=0.00000000000000004492367995028017983415375254583712005 0869800531719*ew^6+0.0000000000000008378484768494891799182306166387981324102296825596 5*ew^5+0.00000000000000005522431406589 9427802789765894852689587104815167352*ew^4+0.0000000000000000 146911611577340785179315785987499964180294993468 49*ew^3+0.000 000000000000011979351185242681597299140874114661346717404726 834*ew^2-0.0000000000000000000839699635623744524665734065 85 698958493547765494 49*ew-0.000000000000000000003723040905315 59502639835221836915576323390966706 94

J[3]=0.000000001396287653570571574662207138195309771989211807
0071*ev+0.000000321736569362350479320642380042960198810751312
25526*ew^5+0.000000332916524777360052071820343090980816286965
94819331*ew^4+0.000000065891103236483021260956200510555934469
425614980251*ew^3-0.000000014502194796324730495895492585449540
4878708307549*ew^2-0.000000025896639470303145276823808149265
0376302930226169440*ew+0.0000000000508191975132258812443532309
13969686213591494118136

J[4]=0.003533742124928111635141433087775284312554356193561 5616*v*ew^4+0.002180171063553251204119523955501421176113052868825*v*ew^3+0.000246834120767041208618495460378545445015627301660 12*v*ew^2-0.000001455465168425825251150185070369471177367962173 0295*v*ew-0.000000075203364912461984107689153015151456273923670 860722*v

J[5]=0.00000000000002469182263912597327197080785436650088580 5230838289*v^2-0.000000000004719021731979233932263858012093117 7672487391351614*ew^5-0.0000000000074233725861437012556372485 08913491275817374285975*ew^4-0.000000000037044087563918384281 802854504031102187085751576865*ew^3-0.000000000005505693794 8011551503818116670012263028105621886299*ew^2+0.000000000000 0289247109748205722101255467392168749848575062251 6*ew-0.0000 0000000000485199500275950858789604761685291061709957821845 75

J[6]=0.000824445295339140004801217492520348576142750871636 34*u*ew^4+0.00154024712861099035299602345961334383656585624 34456*u*ew^3+0.0010186155101412674041097562139745242960275920 262626*u*ew^2+0.0002733014715179344910880975554697358492670 47848435 95*u*ew+0.0000231526498118394429461881625790391710 63101525686 031*u

J[7]=0.155006520853736572782721550488404252690816668614 22*u*v*ew^2+0.096124486745923798453025086330083252761630136 665135*u*v*ew+0.01117811990097039882461133074695487763952818 3070683*u*v

J[8]=0.00000000000002469182263912597327197080785436650088580 5230838289*u^2+0.000000000006618541962195662625184586935775373 8900545315696192*ew^5+0.000000000010411459355962387028179380 989969086880336301691908*ew^4+0.000000000051955227570060962 0984987581216990888395938451519 3*ew^3+0.000000000000077218 685315543857629354030269993964840801627962 3*ew^2-0.000000000 000000405676057886977743474154857219642292365593374 75379*ew-0 .000000000000000023368741490554450106147082788613043426219204 168902

J[9]=0.061970918133163216567183870035153503888001513514567*t*e

```
w^4+0.07786967298515361066019891920879105862334446468288*t*ew^3+0.02952928754574380870962044724506192695086361951332*t*ew^2+0.00321783750214673495126023709862043548276413594934*t*ew+0.00004345685303182307000128985173460828330264279573540*t
J[10]=0.05187413667297784999868866991024835274577850308566*t*v*ew^3+0.03298141656889162536566431212067736852468257601420*t*v*ew^2+0.00424477015171340499438628892165236081282808511703*t*v*ew+0.000058600461674134415884220932579569266784799169692295*t*v
J[11]=0.02179170335426535988026264693729198807682131368572*t*u*ew^3+0.02704103778330774398428661678394321121208245598944*t*u*ew^2+0.00996019886169329319975305970035709497670464004275*t*u*ew+0.00097550648450924773054739454162691886820402290190*t*u
J[12]=0.000000000000002469182263912597327197080785436650088580523083828*t^2-0.0000000000019897101060519652107395125478208499820923810270269*ew^5-0.000000000017025570089756655535494287810528379598119659773602*ew^4-0.00000000000023483592200304199348742150925412718812994059865964*ew^3+0.0000000000019673727812808485015748182293664380763085511692487*ew^2+0.00000000000001978607845702833052592026181105821026866390535766*ew-0.000000000000044219079290739404492269360637167466142801279295227
J[13]=0.05608483317512667868649481132510638542551836978360*s*ew^2+0.03412743926922567589501612244817194784012721403088*s*ew-0.00066282300441698232705078773127382486945402101120119*s-0.04089388502024946597355319193847859030996007939963*t*u*v*ew-0.02901030197176238268232074341665007200071048256574*t*u*v
J[14]=0.02179170335426535988026264693729198807682131368572*s*v*ew-0.00041052862271488237533112188578012581137758918241*s*v-t*u*ew^2-0.91523243331780979308516147067869265402585316644918*t*u*ew-0.18279835739609284049315718860436264109118081404686*t*u
J[15]=0.08268962174313252600656152101084798951526898287462*s*u*ew+0.05187413667297784999868866991024835274577850308566*s*u+t*v*ew^2+0.45088158483424862414602508768019029048144708971225*t*v*ew+0.00681768234810380611795226815290124500741917725349*t*v
J[16]=0.06197091813316321656718387003515350388800151351456*s*u*v+7.54843402550665153673201812730961379367248454713989*t*ew^3+9.29775966257943588582612339280270317033002044556*t*ew^2+2.93333692208165449675224101101090746851954196539922*t*ew+0.0
```

43698233453379147442339681621264612101282386252875*t

J[17]=0.15500652085373657278272155048840425269081666861422*s*t-u*v*ew-0.310069450211982179038598636924866298895502223605*u*v

J[18]=0.000000000000024691822639125973271970807854366500885805230838289*s^2+0.00000000014186637003179206604978647501193712331871881329982*ew^5+0.00000000012139234851393709326934898047082480718477074921104*ew^4+0.0000000000016743805897305182765540212510939603762998421893102*ew^3-0.000000000014027371833250135761788614236060225048185772213358*ew^2-0.0000000000014107477864862906344428789190998451814760012308984*ew-0.00000000000018657929453315597187256856909000195939743454954377

The triangular decomposition to (R.10) is shown in (R.11), which involves five decomposed sets of equations. One decomposed set includes seven entries, into each of which, the seven variables is added one by one, with the order of r, ew, ev, v, u, t, s.

[1]:                                                                       (R.11)
  _[1]=r-1.4
  _[2]=0.082689621743132526006561521010847989515268982874621*ew
      +0.05187413667297784999868866991024835274577850308566
  _[3]=0.00000000013962876535705715746622071381953097719892118070071*ev
      +0.000000000021872894505215127104699976329395744001395648217513
  _[4]=v
  _[5]=0.0000000000000024691822639125973271970807854366500885805230838289*u^2
      -0.00000000000000070387276012025253831506975421079541060390364383305
  _[6]=t
  _[7]=0.0000000000000024691822639125973271970807854366500885805230838289*s^2
        -0.00000000000000050186141759442129979160381879133091558873472043949
[2]:
  _[1]=r-1.4
  _[2]=ew-0.018838757853893431115706795065601087293579573640499
  _[3]=ev-0.018838757853893431115706795065601087293579573640499
  _[4]=v^2-2.03250049593012100694952569568426974696181825868366
  _[5]=u
  _[6]=t
  _[7]=0.0000000000000024691822639125973271970807854366500885805230838289*s^2
        -0.00000000000000050186141759442129979160381879133091558873472043949

[3]:
  _[1]=r-1.4
  _[2]=ew^2+0.620131890042396435807719727384973259779100444721*ew
       +0.072113868754708844917382916961852739039135803532834

_[3]=ev-ew  
_[4]=v^2-14.80995023428197133951905921507729315089177369253934*ew  
    -5.777653339333545581487020716322560290468549512904 11  
_[5]=u^2+2.077131292262844160250221216181327899549542410 21854*ew  
    +0.525266732317117545403256347138142609545167168 24085  
_[6]=t^2+2.077131292262844160250221216181327899549542410 21854*ew  
    +0.525266732317117545403256347138142609545167168 24085  
_[7]=s+8.556707658075617499772076904610507582973719342 31746*t*u*v*ew  
    +6.248349853283175619510255126908401163581602815 53919*t*u*v  

[4]:  
  _[1]=r-1.4  
  _[2]=ew+0.62075494398358690094167914213207711792332315380224  
  _[3]=ev+0.62075494398358690094167914213207711792332315380224  
  _[4]=v  
  _[5]=u^2-0.28506310384917288370291954789447830487339321330884  
  _[6]=t^2-0.28506310384917288370291954789447830487339321330884  
  _[7]=s  

[5]:  
  _[1]=r-1.4  
  _[2]=ew+0.01566503467196175486397478472081382667860308 2481373  
  _[3]=ev+0.62733551780076996606101562740855450717384864 87756  
  _[4]=v^2-2.0325004959301210069495256956842697469618182 5868366  
  _[5]=u  
  _[6]=t^2-0.28506310384917288370291954789447830487339321 330884  
  _[7]=s  

    There are numerical coefficients that are very lengthy ones. This is due to a problem in the algorithm in the Gröbner bases generation [19-21]. The computational procedure applies the Buchberger's algorithm, in which the addition, subtraction, multiplication and division are iterated to the polynomial system. In the intermediate expression through the computation, some polynomials with huge degrees may arise, some of whose coefficients have extreme difference in the numerical scale compared to others. This difference in the scale of coefficients will remain in the final result. To assure the numerical accuracy, we must resort to the computations with the arbitrary precision.

    Though the solutions include complex ones, the physically admissible real solutions are shown in Table. 1. We obtain four combinations, where the two electrons of up or down spins are located the symmetric or asymmetric wavefunctions. This means we obtain both of the ground and the excited states.

### Atomic and electronic structural optimization executed simultaneously

The inter-atomic distance r and the wavefunctions can be optimized at the same time, as is done in Car-Parrinello method. In this case the inter-atomic distance r is an unknown to be determined, not being a fixed constant.

In order to obtain the ground state alone, we add eq. (R.12) into the equations in (R.8).

$$s = v = 0 \tag{R.12}$$

(This is the trick applicable to this example only. In general, the ground state is given as a solution where the sum of the total occupied eigenvalue becomes minimum one. To specify the ground state, it is enough to compute eigenvalues alone. For this purpose, in making the triangular decomposition of the equations, we can prepare the equations including only eigenvalues as unknowns. We have only to solve them.) With this treatment, we can replace the equation to be solved with a simpler one. The part of the equations including r is shown in eq. (R.13), whose real solutions are shown in Table 2. To determine the inter-atomic distance, we have only to solve this part of the equations.

[1]:                                                                                        (R.13)
8942144364*r^23-435341589039*r^22+9813157241157*r^21-134458128500631*r^20+1251986164962728*r^19-8584760758387395*r^18+48176522279858253*r^17-254992901607817871*r^16+1360184656773665254*r^15-6685412705413184235*r^14+26848712421674517351*r^13-82265960807423324641*r^12+185370480318135661708*r^11-295651827763150999108*r^10+307426892321213994312*r^9-148683667595876075980*r^8-97338526988608612178*r^7+245772518836579791529*r^6-200002425723099153061*r^5+47298638179277635737*r^4+46006348188804187952*r^3-41646082527529600720*r^2+13118922400543578496*r-1869747053688110592=0

[2]:
10313892*r^15-376866027*r^14+6245669754*r^13-61144647973*r^12+387764699571*r^11-1646957525797*r^10+4691411679124*r^9-8760215434992*r^8+10281598671237*r^7-7316755042677*r^6+3784010771997*r^5-2194016637700*r^4-299532295668*r^3+1482785614608*r^2-746000940352*r+36100845312=0

[3]:
10313892*r^15-376866027*r^14+6245669754*r^13-61144647973*r^12+378596795571*r^11-1405917509797*r^10+1917133055124*r^9+9044372533008*r^8-55118065080763*r^7+103464030245323*r^6+92432281739997*r^5-770797010005700*r^4+1063674493728332*r^3+652030557238608*r^2-2854269358708352*r+1954998898509312=0

The solutions include the positive and negative real valued ones and the imaginary valued ones. The admissible solutions (r>0) are two in number, as in Table 2. However, the solution which lies in the valid range of the Taylor expansion is only that of r∼1.6. The discrepancy between the solution and the experimental value ( r ∼1.4 ) is due to the numerical error caused by the roughness of the fourth order Taylor expansion and the rationalization of the numerical coefficients, being truncated. In addition, it is also due to the not-optimized orbital exponent in the trial wavefunctions.

### A recipe for an inverse problem

It is demonstrated here how to solve a kind of inverse problem. Suppose a problem, where the energy difference between the occupied and the unoccupied states has a certain value; we should evaluate the inter-atomic distance r at which the energy difference shows this value. This example is a miniature of the inverse problem to find the lattice constants at which the band gap shows the desired width. In this case, we execute RHF (Restricted-Hartree-Fock) calculations. The wavefunctions of the occupied and the unoccupied are given in (R.14) and (R.15). The eigenvalues are denoted as $e_{occ}$, $e_{unocc}$.

$$\phi_{occ}(x) = s(\exp(-x_A) + \exp(-x_B))/\sqrt{\pi} \tag{R.14}$$

$$\phi_{unocc}(x) = t(\exp(-x_A) - \exp(-x_B))/\sqrt{\pi} \tag{R.15}$$

The required equations are presented in (R.16), whose details are omitted here. The set of the equations for the occupied state can be obtained by the same way as was done as the example of UHF calculation. The orthogonality condition to the occupied and the unoccupied states is added to it.

$$
\begin{aligned}
&\frac{\partial \Omega}{\partial s} = 0 \\
&\frac{\partial \Omega}{\partial t} = 0 \\
&\frac{\partial \Omega}{\partial (e_{occ})} = \langle \phi_{occ} | \phi_{occ} \rangle - 1 = 0 \\
&\langle \phi_{unocc} | \phi_{unocc} \rangle = 1 \\
&\langle \phi_{occ} | \phi_{unocc} \rangle = 0 \\
&e_{occ} - e_{unocc} = E_{gap}
\end{aligned}
\tag{R.16}
$$

For example, let us compute r which gives $E_{gap} = e_{unocc} - e_{occ} = 0.9$. The real solutions are those at r=−1.103, 0.307, 1.643, 3.958. The solution in the valid range of the Taylor expansion is only that at r=1.643. The eigenvalues of the occupied and the unoccupied states are shown in Fig.2 as the function of R. It shows this result is the proper one.

Then, in this case, is the structure is stable? If we evaluate the inter-atomic forces, it can be easily judged. On the other hand, with the view of symbolic-numeric solving, we can make use of the following judgment. To do this, in the set of the equations in (R.16) we insert the condition of eq. (R.17), the minimization condition of the energy functional with respect to the inter-atomic distance r.

$$\frac{\partial \Omega}{\partial r} = 0 \tag{R.17}$$

The computed Gröbner base for it is {1} (as a set of polynomials). It includes only a constant polynomial "1". The zeros of the Gröbner bases provide us the solution of the equations. However, the term {1}, as a polynomial, does not become zero. Thus we can conclude that the supposed problem does not have the solution with the stable structure.

### The electronic and atomic structure optimization rewritten in a single matrix eigenvalue problem

We will show another numerical method by means of Stickelberger's theorem. The electronic and atomic structure optimization is rewritten in a single matrix eigenvalue problem. The example is the optimization of inter-atomic distance in $H_2$, where the RHF calculation is executed. The trial wavefunction is given as eq. (R.18).

$$\phi_{occ}(r) = t(\exp(-r_A) + \exp(-r_B))/\sqrt{\pi} \tag{R.18}$$

The eigenvalue is denoted as ev and the inter-atomic distance is r. The HFR equation becomes the set of polynomials expressed by t, ev, and r. This set of equations, in the mathematical sense, constructs a zero-dimensional ideal I in the polynomial ring A=R[t,ev,r], the zeros of which correspond to a residue ring A/I. (R means the rational number field.) A/I is a finite-dimensional vector space over R. Its base is represented by monomials of t, ev, and r, which are shown in (R.19).

$$
\begin{array}{lllll}
b[1]=t*ev*r^3 & b[2]=t*r^4 & b[3]=t^3*ev & b[4]=t*ev^3 & b[5]=t^3*r \\
b[6]=t*ev^2*r & b[7]=t*ev*r^2 & b[8]=t*r^3 & b[9]=ev*r^3 & b[10]=r^4 \\
b[11]=t^3 & b[12]=t^2*ev & b[13]=t*ev^2 & b[14]=ev^3 & b[15]=t^2*r \\
b[16]=t*ev*r & b[17]=ev^2*r & b[18]=t*r^2 & b[19]=ev*r^2 & b[20]=r^3 \\
b[21]=t^2 & b[22]=t*ev & b[23]=ev^2 & b[24]=t*r & b[25]=ev*r \\
b[26]=r^2 & b[27]=t & b[28]=ev & b[29]=r & b[30]=1 \\
\end{array}
\tag{R.19}
$$

The transformation matrix corresponding to the multiplication by the variable t is shown in (R.20).

$$\begin{pmatrix}
0. & 0. & 0. & 0. & 0. & 0. & 0. & 0. & 1. & 0. & 0. & 0. & 0. & 0. & 0. & 0. & 0. & 0. & 0. \\
0. & 0. & 0. & 0. & 0. & 0. & 0. & 0. & 0. & 1. & 0. & 0. & 0. & 0. & 0. & 0. & 0. & 0. & 0. \\
0. & 0. & 0. & 0. & 0. & 0. & 0. & 0. & 0. & 0. & 0. & 1. & 0. & 0. & 0. & 0. & 0. & 0. & 0. \\
0. & 0. & 0. & 0. & 0. & 0. & 0. & 0. & 0. & 0. & 0. & 0. & 1. & 0. & 0. & 0. & 0. & 0. & 0. \\
0. & 0. & 0. & 0. & 0. & 0. & 0. & 0. & 0. & 0. & 0. & 0. & 0. & 1. & 0. & 0. & 0. & 0. & 0. \\
0. & 0. & 0. & 0. & 0. & 0. & 0. & 0. & 0. & 0. & 0. & 0. & 0. & 0. & 0. & 1. & 0. & 0. & 0. \\
0. & 0. & 0. & 0. & 0. & 0. & 0. & 0. & 0. & 0. & 0. & 0. & 0. & 0. & 0. & 0. & 0. & 0. & 1. \\
2.46857 & 0.250358 & 0.000433932 & -0.519297 & 0.000905017 & -0.229989 & 0.400711 & 0.0592776 & 0. & 0.00037686 & 0. & -0.0488928 & 0. & 0.0253457 & 0. & 0.0106224 & 0. \\
2.85865 & 1.8302 & 0.010766 & -3.25489 & 0.00562066 & 0.195796 & 0.755874 & 0.513375 & 0. & -0.0000321824 & 0. & 0.188933 & 0. & 0.177659 & 0. & 0.101971 & 0. \\
-3.28726 & 16.9846 & 0.815632 & -68.3604 & 0.199106 & -6.99935 & 17.2255 & 12.5718 & 0. & 0.010156 & 0. & 2.94616 & 0. & 8.19509 & 0. & 3.31843 & 0. \\
1.10471 & 0.0541875 & 0.0005962 & -0.43266 & 0.000241163 & 0.0161693 & 0.169585 & 0.0191714 & 0. & 0. & 4.36536\times 10^{-6} & 0. & 0.00658837 & 0. & 0.0121792 & 0. & 0.00422577 & 0. \\
-41.1091 & 92.4474 & 0.457252 & -248.171 & 0.949199 & 7.93844 & -9.06291 & 33.1339 & 0. & 0.0550604 & 0. & 7.29159 & 0. & -1.91674 & 0. & 7.47555 & 0. \\
0.381952 & -0.482168 & -0.00366255 & 1.05379 & -0.00214058 & -0.266467 & 0.0658373 & -0.170307 & 0. & -0.0000384651 & 0. & -0.0484318 & 0. & 0.00667468 & 0. & -0.0375156 & 0. \\
-13.5886 & -1.50329 & 0.000693225 & 2.88068 & -0.00965266 & 1.64897 & -2.25783 & -0.319754 & 0. & -0.00357141 & 0. & 0.303898 & 0. & -0.125453 & 0. & -0.05279 & 0. \\
-9.85681 & -9.15049 & -0.0964526 & 22.4732 & -0.0359414 & -1.57568 & -3.20362 & -3.05264 & 0. & 0.00102063 & 0. & -1.41532 & 0. & -1.22295 & 0. & -0.657454 & 0. \\
480.008 & -277.275 & -3.39938 & 331.95 & 0.518864 & -172.803 & 89.1965 & -51.8746 & 0. & 0.670983 & 0. & -36.5975 & 0. & 14.0368 & 0. & 1.13365 & 0. \\
7.34299 & 6.6392 & 0.00938662 & -17.0623 & 0.0147453 & 1.43466 & 1.18689 & 1.56243 & 0. & -0.000518309 & 0. & 0.363976 & 0. & 0.111898 & 0. & 0.278794 & 0. \\
15.2609 & -1.66537 & -0.0325943 & 4.78209 & -0.0255023 & -8.07839 & 2.60648 & -0.91568 & 0. & 0.00589181 & 0. & -0.519365 & 0. & 0.0343565 & 0. & -0.229064 & 0. \\
4.37903 & 18.2502 & 0.31662 & -48.1151 & 0.0770211 & 4.75029 & 3.82373 & 7.43136 & 0. & -0.0068586 & 0. & 3.54526 & 0. & 3.3935 & 0. & 1.7193 & 0. \\
176.488 & 32.4122 & 0.0776154 & -62.9606 & 0.177886 & -12.346 & 26.5689 & 7.04326 & 0. & -0.0170918 & 0. & -5.2778 & 0. & 1.56207 & 0. & 1.18349 & 0. \\
-21.5227 & -39.0633 & -0.494407 & 121.758 & -0.355958 & 0.576694 & -5.67678 & -17.58 & 0. & -0.00940279 & 0. & -3.59933 & 0. & -4.92509 & 0. & -4.18857 & 0. \\
22.4313 & 97.9958 & 1.19706 & -159.224 & 0.120322 & 24.8872 & 3.17625 & 23.7835 & 0. & -0.11519 & 0. & 5.8433 & 0. & 1.73174 & 0. & 2.14857 & 0.
\end{pmatrix}$$

(R.20)

The other transformation matrices are calculated in similar ways. We can evaluate the eigenvector $v_\xi$ of the matrix $m_t$ and can obtain eigenvalues corresponding to variables t and ev by means of the relations in (R.21) and (R.22). The values for t and ev are computed as $\xi_t$ and $\xi_{ev}$. The real valued solutions are identical to those in Table 2, given as the previous example.

$$m_t \cdot v_\xi = \xi_t \cdot v_\xi \tag{R.21}$$

$$m_{ev} \cdot v_\xi = \xi_{ev} \cdot v_\xi \tag{R.22}$$

## DISCUSSION

The advantages by the present method are recapitulated here.

In the present method, the fundamental equation (the HFR equation and the constraints) is expressed as a set of polynomial equations (algebraic molecular orbital equation), combined with each other seamlessly. In the conventional method, the loop for eigenvalue solutions, that of SCF (Self Consistent Field) procedure, and that of the optimization calculation are nested with one another. By contrast, in the present method, those nested loops are unified in a flow of the search for roots in a set of polynomial equations, which afford us clear view in the numerical computation and the shortcut to the needed information. The roots of the triangulated equations are obtained one by one numerically. The eigenvalue solutions for multi-dimensional matrix and the iterative approximation for the mean field, where the difficulty in the convergence generally occurs, are not needed. It is not necessary to iterate the independent optimization toward the atomic structure and the wavefunctions. Concerning with the structural optimization, by means of the elimination of the variables, we can obtain a set of equations which have only atomic coordinates. The roots are the stable optimized structure. It means that the "pseudo" atomic interactions are obtained without SCF calculations. In the optimization of other parameters, there are similar merits. The present method will simulate the cases where the dynamics of the wavefunctions and that of the nuclei are strongly coupled with each other, beyond the adiabatic approximation.

The algebraic molecular orbital equation, expressed as polynomial sets, informs us the relationship among unknown variables. Thus, in order to evaluate those unknowns, we can divide suitable parts of them into the prepared inputs and the expected outputs, respectively. The calculation is not confined to the conventional framework, such as, whose the input is the structure and the output is the electronic the states. The distinction between the forward and the inverse problems are cleared away, and we can treat all of them as the forward problems in a unified way. In order to cope with the inverse problem, we should check whether it is well-posed or not. The present method affords us a key to this. After the transformation from the fundamental equation to the Gröbner bases, the present method can inspect the properness of the problem, i.e. the existence of the solutions. Based on the mathematical ideal theory, it can judge the existence of solution which provides us the zeros of the Gröbner bases. If the solutions exist, it can also be determined whether these are isolated points or the sets with more than one dimension. If the solutions are isolated, the numbers of the solutions, in the range of the real or complex numbers, can be known. (See references on symbolic numeric computations.)

In the computations in previous section, we fixed the orbital exponent z at unity and converted molecular integrals into polynomials of inter-atomic distance R alone. The algebraic molecular equations can be extended to a more general case. We can make multi-variable Taylor expansions for the atomic coordinates and orbital exponents in order to prepare polynomials of those two kinds of variables. If this is done, the equation includes both parameters and we could optimize atomic coordinates and orbital exponents simultaneously.

It should be admitted that the present work lies in a primitive stage. At present, this study does not necessarily afford us sufficiently precise calculation. One reason to this is the constraint by the ability of the hardware. The cost in the symbolic computation sometimes tends to be so massive (especially on memory usage) that we are obliged to use simplified molecular integrals, reduce the degrees of the approximating polynomial and rationalize numerical coefficients in lower accuracies. As for the overall computation cost in the explanatory calculations in the previous section, the generation of the molecular integrals, especially that in two-electron repulsion integrals are the most demanding part. In the desktop pc (2.0GHz dual core CPU, 2.0GB memory), in the case of 1s STO (with the assumption four orbital exponents in the integrand are all equal), the computational time for a two-electron repulsion integral by Mathematica (version 8) reaches the order of 100~1000 seconds. The memory usage for the most complex case (the exchange type two-electron repulsion integral) amounts to 0.5~1GB. If orbital exponents are taken to be different with each other, the computational time amounts to the order of hours, being accompanied with the much increase in the memory usage. (One should note that this is the result by a symbolic computation program which is still under development by the author's research group; there is plenty room for optimization.) The construction of exact and approximate energy functional by Mathematica can be done in a few minutes. As for the symbolic-numeric solving, in the same computational environment, the numerical solution in the examples can be obtained in 1~10 seconds, and the memory usage is about 0.5~1MB, when the computations are processed by Computer algebra system SINGULAR (version 3). This quickness is because of the simplicity of the problem setting. If polynomial equations are much more complicated than those in the above explanatory calculations, for example, with much more lengthy integer numerical coefficients and more number of symbols, the cost on the computational time and memory usage grows so large that the symbolic-numeric computation becomes difficult in a shabby computational environment. In the simple computation of $H_2$ as above, the structural optimization by UHF without artificial symmetrization to wavefunctions results in a severe one. The reinforcement to the computer memory seems to be indispensable to the effective computation.

There is also a fundamental problem in the theoretical side. As can be seen in the starting polynomial equation of (R.8), the scales of the numerical coefficients are almost similar. Meanwhile, after the symbolic computations, in the generated Gröbner bases, the scales of the coefficients show great discrepancies. The extreme growth of coefficients results in the larger computational costs and the lowering of the accuracies in the numerical procedure. Thus, it is one of the major interests in the field of symbolic computation how to avoid such inconveniences and a lot of strategies have been proposed now [19-21]. For example, Brickenstein proposes "slimgb" algorithm in order to keep the intermediate expressions as slim as possible, by regularly replacing a swelled polynomial with a shorter, equivalent one [19]. Lichtblau, based on an empirical study, discusses on important points which should be handled carefully in using approximate arithmetic for coefficients [20]. Arnold presents modular algorithms for the purpose of limiting the enormous growth in rational-numbered coefficients [21]. Those strategies, as well as other ones with the same intention, will be of importance in more complex calculations.

Other difficulties will arise in the application to more complex systems.

Firstly: the generation of the molecular integrals must be burdensome. The shown example, hydrogen molecule is very simple one. We need only one- or two-centered integrals. The molecular integrals are computed from the 1s orbital alone, so that the integrals take the simplest expressions. For a more complex molecule, we must use more general, more complicated atomic basis, which give much more complex molecular integrals. Besides, we have to evaluate three- or four-centered two-electron repulsion integrals. There is a problem: as for the analytic expression of multi-centered repulsion integrals, the derivation by the STO is a formidable one. From this reason, the use of STO has been a minority in quantum chemistry. But, since there are several merits in STO as stated before, the studies to obtain analytic expressions for multi-centered repulsion integrals by means of STO are still pursued [22]. In the present stage, it is more pragmatic to employ GTO, the de facto standard, in which the analytic technique for the derivation of multi-centered integrals has been well established.

Secondly: in the explanatory numerical computations given above, we can evaluate both ground and excited states. Looking from a different angle, this means that we must confront all possible electronic configurations and must single out the necessary one from the solutions. It contrasts to the conventional computations, where it is a trivial work to get a ground state; one has only to compute the necessary number of lower occupied eigenvalues in ascending order from the bottom and put an electron in each of them; in the iterative minimization of the total energy functional, one can gather occupied states only, omitting the calculation of unoccupied states. On the other hand, in the polynomial equations, if one solves them without care, the whole eigenstates (including both of necessary and unnecessary ones, and as many as the Hamiltonian dimension) will appear in the set of solutions indiscriminately. The number of the unnecessary configurations shall grow enormously in more complex systems, where more electrons and a larger basis set are included. In order to sift out the ground states, we can pick up the configuration in which the total sum of the eigenvalues in the occupied states becomes minimal, by firstly solving the equations for energy spectrum obtained by polynomial triangulations. Maybe this tactics will be of use for sorting and indexing excited states. Also, to impose certain point group symmetry on wave function will be effective in the reduction of the computational cost. The RHF structural optimization for hydrogen molecule in the previous section, in which the ground state is extracted from the symmetrized wavefunction, is a clumsy example of this. The switching between symmetrized and asymmentrized wavefunctions leads to the selection of ground or excited states.

Thirdly: the polynomial approximation to the exact functional causes nonsensical solutions, which should be checked with care. In more complicated cases, this situation will be more troublesome. The admissible solutions must lie in a range where the polynomial approximation through the Taylor expansion is quantitatively valid: if the solution appears to be dubious, one must re-examine it by another polynomial approximation with a different center point of the Taylor expansion. One can also use polynomial of higher degrees for this purpose.

In summary, the present work shows that the concept of the "molecular orbital algebraic equation" by means of the "polynomial approximation to molecular integrals" is applicable to the realistic first principles electronic structure calculation, as well as its potentiality to several fundamental problems which are difficult to be handled by the conventional method. At the same time, we must recognize that the hardship to be surmounted is large. However, the improvement in the computer architecture is so rapid that we can expect the achievement of the sufficient accuracy by the present method and its application to complex and large material in future, spurred by the refinement of the symbolic computation theory.

ACKNOWLEDGEMENTS

In this research, the analytic formulas and the Taylor expansions of molecular integrals were generated by Symbolic computation software Mathematica (version 8).

The symbolic-numeric solving was executed by Computer algebra system SINGULAR (version 3).

The author wishes to acknowledge his colleague Dr. J. Yasui for discussions on the concept of molecular orbital algebraic equations. The author also thanks Dr. Yasui for providing him with the symbolic computation software to generate STO molecular integrals.

|  | Solution 1 | Solution 2 | Solution 3 | Solution 4 |
|---|---|---|---|---|
| s (the coefficient for electron 1) | 0.00000 | -1.42566 | 0.00000 | -1.42566 |
| t (the coefficient for electron 1) | -0.53391 | 0.00000 | -0.53391 | 0.00000 |
| u (the coefficient for electron 2) | -0.53391 | -0.53391 | 0.00000 | 0.00000 |
| v (the coefficient for electron 2) | -0.53391 | 0.00000 | -1.42566 | -1.42566 |
| ev (the eigenvalue for electron 1) | -0.62075 | -0.01567 | -0.62734 | 0.01884 |
| ew (the eigenvalue for electron 2) | -0.62075 | -0.62734 | -0.01567 | 0.01884 |
| r (the inter-atomic distance) | 1.40000 | 1.40000 | 1.40000 | 1.40000 |
| The total energy | -1.09624 | -0.49115 | -0.49115 | 0.15503 |
| electron1 | symmetric orbital | asymmetric orbital | symmetric orbital | asymmetric orbital |
| electron2 | symmetric orbital | symmetric orbital | asymmetric orbital | asymmetric orbital |

Table 1. This shows the solutions for equations (R.11). The electron 1 and 2 lie in the up- and down- spin respectively.

|     | Solution1 | Solution2 | Soultion3 |
|-----|-----------|-----------|-----------|
| r   | -1.812    | 1.652     | 6.010     |
| ev  | -6.675    | -0.578    | -17.585   |
| t   | 0.846     | 0.545     | 0.983     |

Table 2. This shows the real solutions in the equations (R.13).

Figure Legends

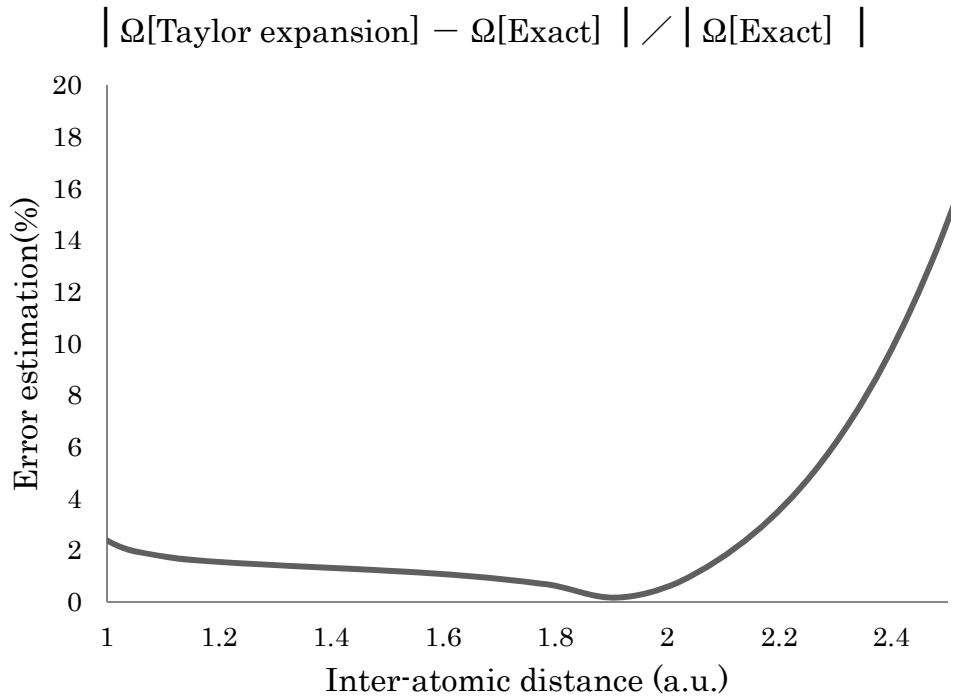

Figure 1: This figure shows the deviation between the exact energy functional $\Omega$ and its polynomial approximation, |Ω(Taylor expansion) - Ω(Exact)|/ |Ω(Exact)| v.s. the interatomic distance r , given in percentage.

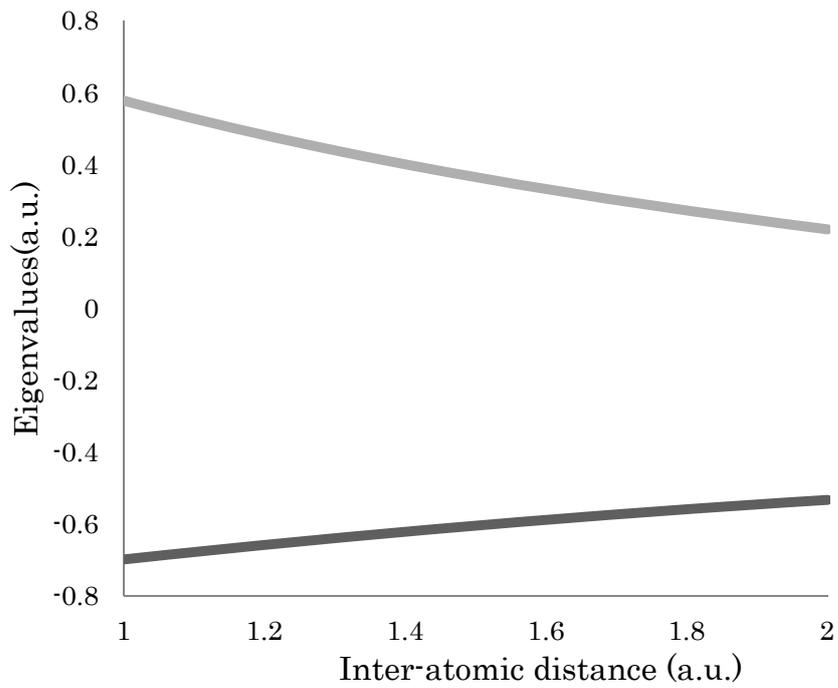

Figure 2: This figure shows the dependence of occupied and unoccupied eigenvalues on the inter-atomic distance. The eigenvalues of the occupied and the unoccupied states are shown by the real and broken lines respectively.

Supplement

1. Molecular Integrals and the polynomial approximation

Molecular orbital calculation requires the matrix elements for orbital overlapping, kinetic energy, electron-nuclei attraction, and electron repulsion. These molecular integrals are listed in below. We follow the notations in the text, where z means the orbital exponents and R is the inter-atomic distance. The localized orbital employed here is of 1s type, defined as $\phi^{1s}(r;z) = \frac{z^{3/2}}{\pi^{1/2}} e^{-zr}$. In general two or four different orbital exponents are necessary in each integral. However, in the present explanatory case, exponents are all equal. All of the required integrals can be obtained from tabulated ones by interchanging the atomic indices; $\phi_A \leftrightarrow \phi_B$ and $r_A \leftrightarrow r_B$. This list is given in Mathematica expressions, where E^x should be read as exp(x), EulerGamma means Euler's Gamma constant, and ExpIntegralEi[x] gives the exponential integral function Ei(z). The integrals are exact ones, except [1s(A)1s(B)|1s(A)1s(B)]. The integral [1s(A)1s(B)|1s(A)1s(B)] is obtained by the truncation of infinite series summation, whose sufficient convergence (below 0.1%) is numerically checked.

Using these integrals, together with LCAO coefficient, we can construct the total energy functional for hydrogen molecule. By means of Taylor expansion at some inter-atomic distance $R_0$, and setting z=1, we can make the polynomial approximation.

In the polynomial approximation, some of numerical coefficients take lengthy expression constructed of rational and irrational numbers. Thus we should re-express coefficients in decimal numbers, truncate them, and express them in rational numbers of moderate sizes, so that we can manipulate them in a modest cost in the symbolic computation.

For the purpose of Taylor expansion, use the "Series[f,{x,x_0,n}]" command in of Mathematica package. In order to simplify the series, use "Expand[f]" and "Simplify[f]" commands. For the purpose of computing and truncating decimal coefficients, apply the numerical evaluation command "N[f]".

For the preparation of polynomial equation, the present work chooses a way in which all numerical coefficients are integer. The approximated energy functional with rational coefficients should be multiplied by a certain integer. (In the present case, an integer, expressed as powers of ten will suffice.) This integer-multiplied energy functional is hereafter processed by SINGULAR package, by which the polynomial equations are generated through symbolic differentiation of the energy functional and numerically solved. In the next section, an example of SINGULAR script is given.

Overlapping:
 $(\phi_A|\phi_A) = 1$.
 $(\phi_A|\phi_B) = (3 + 3*R*z + R^2*z^2)/(3*E^(R*z))$.

Electron-nuclear attraction:

$(\phi_A|\frac{1}{r_A}|\phi_A)$ = z.

$(\phi_A|\frac{1}{r_A}|\phi_B)$ = (-1 + E^(2*R*z) - R*z)/(E^(2*R*z)*R).

$(\phi_A|\frac{1}{r_B}|\phi_A)$ = (z*(1 + R*z))/E^(R*z).

Kinetic energy:
$(\phi_A|-\frac{1}{2}\nabla^2|\phi_A)$ = z^2/2.
$(\phi_A|-\frac{1}{2}\nabla^2|\phi_B)$ = -(z^2*(-3 - 3*R*z + R^2*z^2))/(6*E^(R*z)).

Electron repulsion integrals:

[1s(A)1s(A)|1s(A)1s(A)]= 5z/8.

[1s(A)1s(A)|1s(A)1s(B)] = -(5/(E^(3*R*z)*(16*R))) + 5/(E^(R*z)*(16*R)) -
  ((1/8)*z)/E^(3*R*z) + ((1/8)*z)/E^(R*z) + (R*z^2)/E^(R*z).

[1s(A)1s(A)|1s(B)1s(B)] = 1/R - 1/(E^(2*R*z)*R) - ((11/8)*z)/E^(2*R*z) -
  ((3/4)*R*z^2)/E^(2*R*z) - ((1/6)*R^2*z^3)/E^(2*R*z).

[1s(A)1s(B)|1s(A)1s(B)] = 8474489/(E^(4*R*z)*(20*R)) + 157170489381947/
  (E^(2*R*z)*(600600*R)) - (51498810*EulerGamma)/(E^(2*R*z)*R) +
  21789128416503191090015418875/(E^(4*R*z)*(16*R^22*z^21)) -
  21789128416503191090015418875/(E^(2*R*z)*(16*R^22*z^21)) -
  (982345857947959414659750000*EulerGamma)/(E^(2*R*z)*(R^22*z^21)) +
  607159468933584045545941875/(E^(4*R*z)*(4*R^21*z^20)) +
  9645939037831510179096581255/(E^(2*R*z)*(8*R^21*z^20)) -
  (1964691715895918829319500000*EulerGamma)/(E^(2*R*z)*(R^21*z^20)) +
  402036325851122445680071875/(E^(4*R*z)*(4*R^20*z^19)) +
  1017405755098507245277681875/(E^(2*R*z)*(8*R^20*z^19)) -
  (1917072094564951903266187500*EulerGamma)/(E^(2*R*z)*(R^20*z^19)) +
  481363748933791174444474375/(E^(4*R*z)*(R^19*z^18)) +
  647430285717245054642743125/(E^(2*R*z)*(4*R^19*z^18)) -
  (1214555234602012034106375000*EulerGamma)/(E^(2*R*z)*(R^19*z^18)) +
  283618477831050167468788125/(E^(4*R*z)*(16*R^18*z^17)) +
  1245400769617978428976145625/(E^(2*R*z)*(16*R^18*z^17)) -
  (560875592024342964369000000*EulerGamma)/(E^(2*R*z)*(R^18*z^17)) +
  208773549390763835917188875/(E^(4*R*z)*(4*R^17*z^16)) +
  288310804027404155027155125/(E^(2*R*z)*(8*R^17*z^16)) -
  (200901259120107689243100000*EulerGamma)/(E^(2*R*z)*(R^17*z^16)) +

12638069842759023934170090/(E^(4*R*z)*(R^16*z^15)) +
86085779758080146676036405/(E^(2*R*z)*(8*R^16*z^15)) -
(5798642872500876577479300*EulerGamma)/(E^(2*R*z)*(R^16*z^15)) +
10260099869607253713699765/(E^(4*R*z)*(4*R^15*z^14)) +
59528364307056868888280915/(E^(2*R*z)*(2*R^15*z^14)) -
(13829399989978108121586000*EulerGamma)/(E^(2*R*z)*(R^15*z^14)) +
17674852664304583011485/(E^(4*R*z)*(4*R^14*z^13)) +
24647349178235374531652225/(E^(2*R*z)*(4*R^14*z^13)) -
(27723467756959514139600000*EulerGamma)/(E^(2*R*z)*(R^14*z^13)) +
260546594972182484540855/(E^(4*R*z)*(4*R^13*z^12)) +
46550409944080458814843585/(E^(2*R*z)*(4*R^13*z^12)) -
(4726564104366340956720000*EulerGamma)/(E^(2*R*z)*(R^13*z^12)) +
825672402044545165155/(E^(4*R*z)*(R^12*z^11)) +
17567010500967907906365/(E^(2*R*z)*(R^12*z^11)) -
(6906465270665122862700*EulerGamma)/(E^(2*R*z)*(R^12*z^11)) +
90190156108073766507/(E^(4*R*z)*(R^11*z^10)) +
2391179995126391390172/(E^(2*R*z)*(R^11*z^10)) -
(868869886740164389080*EulerGamma)/(E^(2*R*z)*(R^11*z^10)) +
135745441000845592563/(E^(4*R*z)*(16*R^10*z^9)) +
4296572552616721868259/(E^(2*R*z)*(16*R^10*z^9)) -
(94278742226841332160*EulerGamma)/(E^(2*R*z)*(R^10*z^9)) +
2739501788182716213/(E^(4*R*z)*(4*R^9*z^8)) +
214589308045422019635/(E^(2*R*z)*(8*R^9*z^8)) -
(8815655375146620000*EulerGamma)/(E^(2*R*z)*(R^9*z^8)) +
94245802656830829/(E^(4*R*z)*(2*R^8*z^7)) +
17893664377186069179/(E^(2*R*z)*(8*R^8*z^7)) -
(707774737523473620*EulerGamma)/(E^(2*R*z)*(R^8*z^7)) +
5468299153048413/(E^(4*R*z)*(2*R^7*z^6)) +
650526570598061457/(E^(2*R*z)*(4*R^7*z^6)) -
(48458799864394920*EulerGamma)/(E^(2*R*z)*(R^7*z^6)) +
10533375099861129/(E^(4*R*z)*(80*R^6*z^5)) +
1099218032465226501/(E^(2*R*z)*(112*R^6*z^5)) -
(2798977418996616*EulerGamma)/(E^(2*R*z)*(R^6*z^5)) +
102770483391159/(E^(4*R*z)*(20*R^5*z^4)) +
27993723485927565/(E^(2*R*z)*(56*R^5*z^4)) -
(134197512566832*EulerGamma)/(E^(2*R*z)*(R^5*z^4)) +
3128067354891/(E^(4*R*z)*(20*R^4*z^3)) + 1148458168092019/
 (E^(2*R*z)*(56*R^4*z^3)) - (5213361064230*EulerGamma)/
 (E^(2*R*z)*(R^4*z^3)) + 17301023849/(E^(4*R*z)*(5*R^3*z^2)) +
186823733331981/(E^(2*R*z)*(28*R^3*z^2)) - (158129503884*EulerGamma)/
 (E^(2*R*z)*(R^3*z^2)) + 4104003633/(E^(4*R*z)*(80*R^2*z)) +
1800550441175/(E^(2*R*z)*(112*R^2*z)) - (3523760064*EulerGamma)/

$$\begin{aligned}
&(E^{\wedge}(2*R*z)*(R^{\wedge}2*z)) + ((5/3)*z)/E^{\wedge}(4*R*z) + \\
&(2042740211921*z)/(E^{\wedge}(2*R*z)*900900) - (372680*\text{EulerGamma}*z)/ \\
&\quad E^{\wedge}(2*R*z) + ((11/15)*R*z^{\wedge}2)/E^{\wedge}(4*R*z) + (244445819*R*z^{\wedge}2)/ \\
&\quad (E^{\wedge}(2*R*z)*90090) + ((2/15)*R^{\wedge}2*z^{\wedge}3)/E^{\wedge}(4*R*z) - \\
&(4493872*R^{\wedge}2*z^{\wedge}3)/(E^{\wedge}(2*R*z)*225225) + (986603*R^{\wedge}3*z^{\wedge}4)/ \\
&\quad (E^{\wedge}(2*R*z)*1351350) - (51498810*E^{\wedge}(2*R*z)*\text{ExpIntegralEi}[-4*R*z])/R - \\
&(19646917158959188293195000*\text{ExpIntegralEi}[-4*R*z])/(R^{\wedge}22*z^{\wedge}21) + \\
&(9823458579479594146597500*E^{\wedge}(2*R*z)*\text{ExpIntegralEi}[-4*R*z])/ \\
&\quad (R^{\wedge}22*z^{\wedge}21) - (19646917158959188293195000*E^{\wedge}(2*R*z)* \\
&\quad \text{ExpIntegralEi}[-4*R*z])/(R^{\wedge}21*z^{\wedge}20) + \\
&(9523924266193385210662500*\text{ExpIntegralEi}[-4*R*z])/(R^{\wedge}20*z^{\wedge}19) + \\
&(19170720945649519032661875 0*E^{\wedge}(2*R*z)*\text{ExpIntegralEi}[-4*R*z])/ \\
&\quad (R^{\wedge}20*z^{\wedge}19) - (121455523460201203410637500*E^{\wedge}(2*R*z)* \\
&\quad \text{ExpIntegralEi}[-4*R*z])/(R^{\wedge}19*z^{\wedge}18) - \\
&(243519210860774673825000*\text{ExpIntegralEi}[-4*R*z])/(R^{\wedge}18*z^{\wedge}17) + \\
&(560875592024342964369000 00*E^{\wedge}(2*R*z)*\text{ExpIntegralEi}[-4*R*z])/ \\
&\quad (R^{\wedge}18*z^{\wedge}17) - (20090125912010768924310000*E^{\wedge}(2*R*z)* \\
&\quad \text{ExpIntegralEi}[-4*R*z])/(R^{\wedge}17*z^{\wedge}16) + \\
&(4396196110151201756400*\text{ExpIntegralEi}[-4*R*z])/(R^{\wedge}16*z^{\wedge}15) + \\
&(5798642872500876577479300*E^{\wedge}(2*R*z)*\text{ExpIntegralEi}[-4*R*z])/ \\
&\quad (R^{\wedge}16*z^{\wedge}15) - (1382939998997810812158600*E^{\wedge}(2*R*z)* \\
&\quad \text{ExpIntegralEi}[-4*R*z])/(R^{\wedge}15*z^{\wedge}14) - \\
&(63340989550903864800*\text{ExpIntegralEi}[-4*R*z])/(R^{\wedge}14*z^{\wedge}13) + \\
&(277234677569595141396000*E^{\wedge}(2*R*z)*\text{ExpIntegralEi}[-4*R*z])/ \\
&\quad (R^{\wedge}14*z^{\wedge}13) - (47265641043663409567200*E^{\wedge}(2*R*z)* \\
&\quad \text{ExpIntegralEi}[-4*R*z])/(R^{\wedge}13*z^{\wedge}12) + \\
&(781761906312987600*\text{ExpIntegralEi}[-4*R*z])/(R^{\wedge}12*z^{\wedge}11) + \\
&(6906465270665122862700*E^{\wedge}(2*R*z)*\text{ExpIntegralEi}[-4*R*z])/(R^{\wedge}12*z^{\wedge}11) - \\
&(868869886740164389080*E^{\wedge}(2*R*z)*\text{ExpIntegralEi}[-4*R*z])/(R^{\wedge}11*z^{\wedge}10) - \\
&(8683305154524000*\text{ExpIntegralEi}[-4*R*z])/(R^{\wedge}10*z^{\wedge}9) + \\
&(94278742226841332160*E^{\wedge}(2*R*z)*\text{ExpIntegralEi}[-4*R*z])/(R^{\wedge}10*z^{\wedge}9) - \\
&(8815655375146620000*E^{\wedge}(2*R*z)*\text{ExpIntegralEi}[-4*R*z])/(R^{\wedge}9*z^{\wedge}8) + \\
&(90453720357000*\text{ExpIntegralEi}[-4*R*z])/(R^{\wedge}8*z^{\wedge}7) + \\
&(707774737523473620*E^{\wedge}(2*R*z)*\text{ExpIntegralEi}[-4*R*z])/(R^{\wedge}8*z^{\wedge}7) - \\
&(48458799864394920*E^{\wedge}(2*R*z)*\text{ExpIntegralEi}[-4*R*z])/(R^{\wedge}7*z^{\wedge}6) - \\
&(923655230784*\text{ExpIntegralEi}[-4*R*z])/(R^{\wedge}6*z^{\wedge}5) + \\
&(2798977418996616*E^{\wedge}(2*R*z)*\text{ExpIntegralEi}[-4*R*z])/(R^{\wedge}6*z^{\wedge}5) - \\
&(134197512566832*E^{\wedge}(2*R*z)*\text{ExpIntegralEi}[-4*R*z])/(R^{\wedge}5*z^{\wedge}4) + \\
&(9947723148*\text{ExpIntegralEi}[-4*R*z])/(R^{\wedge}4*z^{\wedge}3) + \\
&(5213361064230*E^{\wedge}(2*R*z)*\text{ExpIntegralEi}[-4*R*z])/(R^{\wedge}4*z^{\wedge}3) - \\
&(158129503884*E^{\wedge}(2*R*z)*\text{ExpIntegralEi}[-4*R*z])/(R^{\wedge}3*z^{\wedge}2) - \\
&(152074296*\text{ExpIntegralEi}[-4*R*z])/(R^{\wedge}2*z) +
\end{aligned}$$

(3523760064*E^(2*R*z)*ExpIntegralEi[-4*R*z])/(R^2*z) +
204160*z*ExpIntegralEi[-4*R*z] + 372680*E^(2*R*z)*z*
 ExpIntegralEi[-4*R*z] - (12*ExpIntegralEi[-2*R*z])/(5*R) +
(514988112*ExpIntegralEi[-2*R*z])/(E^(2*R*z)*(5*R)) +
(19646917158959188293195000*ExpIntegralEi[-2*R*z])/
 (E^(2*R*z)*(R^22*z^21)) + (39293834317918376586390000*
  ExpIntegralEi[-2*R*z])/(E^(2*R*z)*(R^21*z^20)) +
(38341441891299038065323750*ExpIntegralEi[-2*R*z])/
 (E^(2*R*z)*(R^20*z^19)) + (24291104692040240682127500*
  ExpIntegralEi[-2*R*z])/(E^(2*R*z)*(R^19*z^18)) +
(11217511840486859287380000*ExpIntegralEi[-2*R*z])/
 (E^(2*R*z)*(R^18*z^17)) + (4018025182402153784862000*
  ExpIntegralEi[-2*R*z])/(E^(2*R*z)*(R^17*z^16)) +
(1159728574500175315495860*ExpIntegralEi[-2*R*z])/
 (E^(2*R*z)*(R^16*z^15)) + (276587999799562162431720*
  ExpIntegralEi[-2*R*z])/(E^(2*R*z)*(R^15*z^14)) +
(55446935513919028279200*ExpIntegralEi[-2*R*z])/
 (E^(2*R*z)*(R^14*z^13)) + (9453128208732681913440*
  ExpIntegralEi[-2*R*z])/(E^(2*R*z)*(R^13*z^12)) +
(1381293054133024572540*ExpIntegralEi[-2*R*z])/
 (E^(2*R*z)*(R^12*z^11)) + (173773977348032877816*
  ExpIntegralEi[-2*R*z])/(E^(2*R*z)*(R^11*z^10)) +
(18855748445368266432*ExpIntegralEi[-2*R*z])/(E^(2*R*z)*(R^10*z^9)) +
(1763131075029324000*ExpIntegralEi[-2*R*z])/(E^(2*R*z)*(R^9*z^8)) +
(141554947504694724*ExpIntegralEi[-2*R*z])/(E^(2*R*z)*(R^8*z^7)) +
(9691759972878984*ExpIntegralEi[-2*R*z])/(E^(2*R*z)*(R^7*z^6)) +
(559795483799323*ExpIntegralEi[-2*R*z])/(E^(2*R*z)*(R^6*z^5)) +
(26839502513366*ExpIntegralEi[-2*R*z])/(E^(2*R*z)*(R^5*z^4)) +
(1042672212846*ExpIntegralEi[-2*R*z])/(E^(2*R*z)*(R^4*z^3)) +
(31625900776*ExpIntegralEi[-2*R*z])/(E^(2*R*z)*(R^3*z^2)) +
(7047520128*ExpIntegralEi[-2*R*z])/(E^(2*R*z)*(R^2*z)) +
((3726824/5)*z*ExpIntegralEi[-2*R*z])/E^(2*R*z) +
(4/5)*R*z^2*ExpIntegralEi[-2*R*z] + (4*R*z^2*ExpIntegralEi[-2*R*z])/
 E^(2*R*z) + ((8/5)*R^2*z^3*ExpIntegralEi[-2*R*z])/E^(2*R*z) -
(4/15)*R^3*z^4*ExpIntegralEi[-2*R*z] +
((4/15)*R^3*z^4*ExpIntegralEi[-2*R*z])/E^(2*R*z) -
(51498810*Log[R])/(E^(2*R*z)*R) - (9823458579479594146597500*Log[R])/
 (E^(2*R*z)*(R^22*z^21)) - (19646917158959188293195000*Log[R])/
 (E^(2*R*z)*(R^21*z^20)) - (19170720945649519032661875*Log[R])/
 (E^(2*R*z)*(R^20*z^19)) - (12145552346020120341063750*Log[R])/
 (E^(2*R*z)*(R^19*z^18)) - (5608755920243429643690000*Log[R])/
 (E^(2*R*z)*(R^18*z^17)) - (2009012591201076892431000*Log[R])/

$$
\begin{aligned}
&(E^{\wedge}(2*R*z)*(R^{\wedge}17*z^{\wedge}16)) - (5798642872500876577479300*\text{Log}[R])/\\
&(E^{\wedge}(2*R*z)*(R^{\wedge}16*z^{\wedge}15)) - (13829399989978108121586 00*\text{Log}[R])/\\
&(E^{\wedge}(2*R*z)*(R^{\wedge}15*z^{\wedge}14)) - (277234677569595141396000*\text{Log}[R])/\\
&(E^{\wedge}(2*R*z)*(R^{\wedge}14*z^{\wedge}13)) - (47265641043663409567200*\text{Log}[R])/\\
&(E^{\wedge}(2*R*z)*(R^{\wedge}13*z^{\wedge}12)) - (6906465270665122862700*\text{Log}[R])/\\
&(E^{\wedge}(2*R*z)*(R^{\wedge}12*z^{\wedge}11)) - (868869886740164389080*\text{Log}[R])/\\
&(E^{\wedge}(2*R*z)*(R^{\wedge}11*z^{\wedge}10)) - (94278742226841332160*\text{Log}[R])/\\
&(E^{\wedge}(2*R*z)*(R^{\wedge}10*z^{\wedge}9)) - (8815655375146620000*\text{Log}[R])/\\
&(E^{\wedge}(2*R*z)*(R^{\wedge}9*z^{\wedge}8)) - (707774737523473620*\text{Log}[R])/\\
&(E^{\wedge}(2*R*z)*(R^{\wedge}8*z^{\wedge}7)) - (48458799864394920*\text{Log}[R])/\\
&(E^{\wedge}(2*R*z)*(R^{\wedge}7*z^{\wedge}6)) - (2798977418996616*\text{Log}[R])/\\
&(E^{\wedge}(2*R*z)*(R^{\wedge}6*z^{\wedge}5)) - (134197512566832*\text{Log}[R])/\\
&(E^{\wedge}(2*R*z)*(R^{\wedge}5*z^{\wedge}4)) - (5213361064230*\text{Log}[R])/\\
&(E^{\wedge}(2*R*z)*(R^{\wedge}4*z^{\wedge}3)) - (158129503884*\text{Log}[R])/\\
&(E^{\wedge}(2*R*z)*(R^{\wedge}3*z^{\wedge}2)) - (3523760064*\text{Log}[R])/(E^{\wedge}(2*R*z)*(R^{\wedge}2*z)) -\\
&(372680*z*\text{Log}[R])/E^{\wedge}(2*R*z) + (51498810*\text{Log}[-(1/(2*z))])/\\
&(E^{\wedge}(2*R*z)*R) + (9823458579479594 1465975000*\text{Log}[-(1/(2*z))])/\\
&(E^{\wedge}(2*R*z)*(R^{\wedge}22*z^{\wedge}21)) + (19646917158959188293 1950000*\\
&\quad\text{Log}[-(1/(2*z))])/(E^{\wedge}(2*R*z)*(R^{\wedge}21*z^{\wedge}20)) +\\
&(19170720945649519032661875 0*\text{Log}[-(1/(2*z))])/\\
&(E^{\wedge}(2*R*z)*(R^{\wedge}20*z^{\wedge}19)) + (121455523460201203410637500*\\
&\quad\text{Log}[-(1/(2*z))])/(E^{\wedge}(2*R*z)*(R^{\wedge}19*z^{\wedge}18)) +\\
&(56087559202434296436900000*\text{Log}[-(1/(2*z))])/(E^{\wedge}(2*R*z)*(R^{\wedge}18*z^{\wedge}17)) +\\
&(20090125912010768924310000*\text{Log}[-(1/(2*z))])/(E^{\wedge}(2*R*z)*(R^{\wedge}17*z^{\wedge}16)) +\\
&(5798642872500876577479300*\text{Log}[-(1/(2*z))])/(E^{\wedge}(2*R*z)*(R^{\wedge}16*z^{\wedge}15)) +\\
&(1382939998997810812158600*\text{Log}[-(1/(2*z))])/(E^{\wedge}(2*R*z)*(R^{\wedge}15*z^{\wedge}14)) +\\
&(277234677569595141396000*\text{Log}[-(1/(2*z))])/(E^{\wedge}(2*R*z)*(R^{\wedge}14*z^{\wedge}13)) +\\
&(47265641043663409567200*\text{Log}[-(1/(2*z))])/(E^{\wedge}(2*R*z)*(R^{\wedge}13*z^{\wedge}12)) +\\
&(6906465270665122862700*\text{Log}[-(1/(2*z))])/(E^{\wedge}(2*R*z)*(R^{\wedge}12*z^{\wedge}11)) +\\
&(868869886740164389080*\text{Log}[-(1/(2*z))])/(E^{\wedge}(2*R*z)*(R^{\wedge}11*z^{\wedge}10)) +\\
&(94278742226841332160*\text{Log}[-(1/(2*z))])/(E^{\wedge}(2*R*z)*(R^{\wedge}10*z^{\wedge}9)) +\\
&(8815655375146620000*\text{Log}[-(1/(2*z))])/(E^{\wedge}(2*R*z)*(R^{\wedge}9*z^{\wedge}8)) +\\
&(707774737523473620*\text{Log}[-(1/(2*z))])/(E^{\wedge}(2*R*z)*(R^{\wedge}8*z^{\wedge}7)) +\\
&(48458799864394920*\text{Log}[-(1/(2*z))])/(E^{\wedge}(2*R*z)*(R^{\wedge}7*z^{\wedge}6)) +\\
&(2798977418996616*\text{Log}[-(1/(2*z))])/(E^{\wedge}(2*R*z)*(R^{\wedge}6*z^{\wedge}5)) +\\
&(134197512566832*\text{Log}[-(1/(2*z))])/(E^{\wedge}(2*R*z)*(R^{\wedge}5*z^{\wedge}4)) +\\
&(5213361064230*\text{Log}[-(1/(2*z))])/(E^{\wedge}(2*R*z)*(R^{\wedge}4*z^{\wedge}3)) +\\
&(158129503884*\text{Log}[-(1/(2*z))])/(E^{\wedge}(2*R*z)*(R^{\wedge}3*z^{\wedge}2)) +\\
&(3523760064*\text{Log}[-(1/(2*z))])/(E^{\wedge}(2*R*z)*(R^{\wedge}2*z)) +\\
&(372680*z*\text{Log}[-(1/(2*z))])/E^{\wedge}(2*R*z) - (25749405*\text{Log}[-(1/(4*z))])/\\
&(E^{\wedge}(2*R*z)*R) - (4911729289737979 0732987500*\text{Log}[-(1/(4*z))])/\\
&(E^{\wedge}(2*R*z)*(R^{\wedge}22*z^{\wedge}21)) - (9823458579479594 1465975000*\\
\end{aligned}
$$

Log[-(1/(4*z))])/(E^(2*R*z)*(R^21*z^20)) - 
(9585360472824759516330975*Log[-(1/(4*z))])/(E^(2*R*z)*(R^20*z^19)) - 
(60727761730100601705318750*Log[-(1/(4*z))])/(E^(2*R*z)*(R^19*z^18)) - 
(280437796012171482184500000*Log[-(1/(4*z))])/(E^(2*R*z)*(R^18*z^17)) - 
(1004506295600538446215000000*Log[-(1/(4*z))])/(E^(2*R*z)*(R^17*z^16)) - 
(2899321436250438288739650*Log[-(1/(4*z))])/(E^(2*R*z)*(R^16*z^15)) - 
(6914699949890540679300*Log[-(1/(4*z))])/(E^(2*R*z)*(R^15*z^14)) - 
(13861733878479757069800*Log[-(1/(4*z))])/(E^(2*R*z)*(R^14*z^13)) - 
(23632820521831704783600*Log[-(1/(4*z))])/(E^(2*R*z)*(R^13*z^12)) - 
(34532326353325614331350*Log[-(1/(4*z))])/(E^(2*R*z)*(R^12*z^11)) - 
(43443494337008219454*Log[-(1/(4*z))])/(E^(2*R*z)*(R^11*z^10)) - 
(47139371113420666080*Log[-(1/(4*z))])/(E^(2*R*z)*(R^10*z^9)) - 
(44078276875733100*Log[-(1/(4*z))])/(E^(2*R*z)*(R^9*z^8)) - 
(35388736876173681*Log[-(1/(4*z))])/(E^(2*R*z)*(R^8*z^7)) - 
(24229399932197460*Log[-(1/(4*z))])/(E^(2*R*z)*(R^7*z^6)) - 
(1399488709498308*Log[-(1/(4*z))])/(E^(2*R*z)*(R^6*z^5)) - 
(67098756283416*Log[-(1/(4*z))])/(E^(2*R*z)*(R^5*z^4)) - 
(2606680532115*Log[-(1/(4*z))])/(E^(2*R*z)*(R^4*z^3)) - 
(79064751942*Log[-(1/(4*z))])/(E^(2*R*z)*(R^3*z^2)) - 
(1761880032*Log[-(1/(4*z))])/(E^(2*R*z)*(R^2*z)) - 
(186340*z*Log[-(1/(4*z))])/E^(2*R*z) + (25749405*Log[-4*z])/
 (E^(2*R*z)*R) + (4911729289739797072987500*Log[-4*z])/
 (E^(2*R*z)*(R^22*z^21)) + (98234585794795941465975000*Log[-4*z])/
 (E^(2*R*z)*(R^21*z^20)) + (9585360472824759516330975*Log[-4*z])/
 (E^(2*R*z)*(R^20*z^19)) + (60727761730100601705318750*Log[-4*z])/
 (E^(2*R*z)*(R^19*z^18)) + (280437796012171482184500000*Log[-4*z])/
 (E^(2*R*z)*(R^18*z^17)) + (1004506295600538446215000000*Log[-4*z])/
 (E^(2*R*z)*(R^17*z^16)) + (2899321436250438288739650*Log[-4*z])/
 (E^(2*R*z)*(R^16*z^15)) + (6914699949890540679300*Log[-4*z])/
 (E^(2*R*z)*(R^15*z^14)) + (13861733878479757069800*Log[-4*z])/
 (E^(2*R*z)*(R^14*z^13)) + (23632820521831704783600*Log[-4*z])/
 (E^(2*R*z)*(R^13*z^12)) + (34532326353325614331350*Log[-4*z])/
 (E^(2*R*z)*(R^12*z^11)) + (43443494337008219454*Log[-4*z])/
 (E^(2*R*z)*(R^11*z^10)) + (47139371113420666080*Log[-4*z])/
 (E^(2*R*z)*(R^10*z^9)) + (44078276875733100*Log[-4*z])/
 (E^(2*R*z)*(R^9*z^8)) + (35388736876173681*Log[-4*z])/
 (E^(2*R*z)*(R^8*z^7)) + (24229399932197460*Log[-4*z])/
 (E^(2*R*z)*(R^7*z^6)) + (1399488709498308*Log[-4*z])/
 (E^(2*R*z)*(R^6*z^5)) + (67098756283416*Log[-4*z])/
 (E^(2*R*z)*(R^5*z^4)) + (2606680532115*Log[-4*z])/
 (E^(2*R*z)*(R^4*z^3)) + (79064751942*Log[-4*z])/
 (E^(2*R*z)*(R^3*z^2)) + (1761880032*Log[-4*z])/

$(E^{\wedge}(2*R*z)*(R^{\wedge}2*z)) + (186340*z*\text{Log}[-4*z])/E^{\wedge}(2*R*z) - (51498810*\text{Log}[-2*z])/(E^{\wedge}(2*R*z)*R) - (982345857947959414659750000*\text{Log}[-2*z])/(E^{\wedge}(2*R*z)*(R^{\wedge}22*z^{\wedge}21)) - (19646917158959188293195000*\text{Log}[-2*z])/(E^{\wedge}(2*R*z)*(R^{\wedge}21*z^{\wedge}20)) - (191707209456495190326618750*\text{Log}[-2*z])/(E^{\wedge}(2*R*z)*(R^{\wedge}20*z^{\wedge}19)) - (1214555234602012034106375000*\text{Log}[-2*z])/(E^{\wedge}(2*R*z)*(R^{\wedge}19*z^{\wedge}18)) - (5608755920243429643690000*\text{Log}[-2*z])/(E^{\wedge}(2*R*z)*(R^{\wedge}18*z^{\wedge}17)) - (20090125912010768924310000*\text{Log}[-2*z])/(E^{\wedge}(2*R*z)*(R^{\wedge}17*z^{\wedge}16)) - (57986428725008765774793000*\text{Log}[-2*z])/(E^{\wedge}(2*R*z)*(R^{\wedge}16*z^{\wedge}15)) - (138293999899781081215860000*\text{Log}[-2*z])/(E^{\wedge}(2*R*z)*(R^{\wedge}15*z^{\wedge}14)) - (277234677569595141396000*\text{Log}[-2*z])/(E^{\wedge}(2*R*z)*(R^{\wedge}14*z^{\wedge}13)) - (472656410436634095672000*\text{Log}[-2*z])/(E^{\wedge}(2*R*z)*(R^{\wedge}13*z^{\wedge}12)) - (690646527066512286270000*\text{Log}[-2*z])/(E^{\wedge}(2*R*z)*(R^{\wedge}12*z^{\wedge}11)) - (868869886740164389080*\text{Log}[-2*z])/(E^{\wedge}(2*R*z)*(R^{\wedge}11*z^{\wedge}10)) - (942787422268413321600*\text{Log}[-2*z])/(E^{\wedge}(2*R*z)*(R^{\wedge}10*z^{\wedge}9)) - (88156553751466200000*\text{Log}[-2*z])/(E^{\wedge}(2*R*z)*(R^{\wedge}9*z^{\wedge}8)) - (707774737523473620*\text{Log}[-2*z])/(E^{\wedge}(2*R*z)*(R^{\wedge}8*z^{\wedge}7)) - (48458799864394920*\text{Log}[-2*z])/(E^{\wedge}(2*R*z)*(R^{\wedge}7*z^{\wedge}6)) - (2798977418996616*\text{Log}[-2*z])/(E^{\wedge}(2*R*z)*(R^{\wedge}6*z^{\wedge}5)) - (134197512566832*\text{Log}[-2*z])/(E^{\wedge}(2*R*z)*(R^{\wedge}5*z^{\wedge}4)) - (5213361064230*\text{Log}[-2*z])/(E^{\wedge}(2*R*z)*(R^{\wedge}4*z^{\wedge}3)) - (158129503884*\text{Log}[-2*z])/(E^{\wedge}(2*R*z)*(R^{\wedge}3*z^{\wedge}2)) - (3523760064*\text{Log}[-2*z])/(E^{\wedge}(2*R*z)*(R^{\wedge}2*z)) - (372680*z*\text{Log}[-2*z])/E^{\wedge}(2*R*z)$.

The nuclear-nuclear repulsion: $1/R$.

2. An example for symbolic numeric solving script

The approximated polynomial energy functional can be processed by a computational algebra system SINGULAR. Several commands in SINGULAR are applied for this purpose. For the symbolic differentiation, use "diff(…)" command. For the generation of Gröbner basis, use "std(…)" command. The option "redSB" for the computation of reduced Gröbner bases must be specified as "option (redSB)", because the reduced bases are necessary hereafter. One can try "slimgb(…)" command in order to repress the swell in the coefficients in Gröbner bases generation. For the purpose of Gröbner bases triangulation, use "triangM(…)". (Different algorithms for triangulation are also available in SIGULAR.) For the numerical solution, use "laguerre_solve(…)" and "triang_solve(…)". For the numerical solution by means of Stickelberger's theorem, one should use "qbase(…)" and "matmult(…)" commands to generate the monomial bases in the quotient ring and the necessary transformation matrices.

An example of SINGULAR script for the symbolic numeric solving is given here. This script is used for the electronic structure calculation for $H_2$ with fixed atomic distance (r=7/5). The Gröbner bases are generated with the degree reverse lexicographical ordering, converted into those with lexicographical ordering by FGLM (Faugere, Gianni, Lazard, and Mora) algorithm and then decomposed into the triangular sets. (The change in the monomial ordering by FGLM is a speed-up technique for Gröbner bases computation.) The computation starts with the ground field of rational numbers to compute the Gröbner bases with integer coefficients. On the way, the ground field is changed into that of real floating point numbers with precision of 50 digits and the floating number solutions are computed in the end. This script uses "triang.lib" in SINGULAR package for triangulation of Gröbner bases and "solve.lib" for numerical solving of polynomial equations. The meaning for the variables is same as that in the text. For the details of SINGULAR commands, one should consult with the users' manual of SINGULAR.

```
//Computation of (a,b,c, d,r,ev,ew).
// Energy functional: Taylor expansion at r=7/5,4th degree.
// (a,b): LCAO coefficient for eigenvalue ev.
// (c,d): LCAO coefficient for eigenvalue ew.
// r:inter-atomic distance.
// p0 = Omega(Energy functional)*1000. (Only integral coefficients are involved.)
// p1 = d(Omega)/da :Derivatives.
// p2 = d(Omega)/db.
// p3 = d(Omega)/dc.
// p4 = d(Omega)/dd.
// p5 = d(Omega)/dr.
// p6 = d(Omega)/d(ev) :normalization.
// p7 = d(Omega)/d(ew) :normalization.
// On the way, we make the following change of variables:
// a=t+s; b=t-s;c=u+v; d=u-v;
//For the solution at r=7/5, the definition of the set of polynomial equations
// is expressed by an ideal.
//ideal i1=p1,p2,p3,p4,p5,p6,5*r-7;
//For structural optimization, the ideal is given as;
//ideal i1=p1,p2,p3,p4,p5,p6,p7;
//If one wants to symmetrize or asymmetrize wavefunction,
//add any of "s","v","t","u". The example for the symmetrized case given as:
// ideal i1=p1,p2,p3,p4,p5,p6,p7,s,v;

LIB "triang.lib";
```

```
LIB "solve.lib";
// Ring definition.
ring R=0,(s,t,u,v,a,b,c,d,ev,ew,r),lp;
//This definition declares a polynomial ring called R with a ground field of
//characteristic 0 (i.e., the rational numbers) and ring variables
//(s,t,u,v,a,b,c,d,ev,ew,r) . The "lp" at the end determines that the lexicographical
//ordering ordering will be used.
//Instead, one can define the ring in another way: for example,
// ring R=(real, 50),(s,t,u,v,a,b,c,d,ev,ew,r),lp;
//In this definition, the ground field is that of floating point numbers with precision
//extended to 50 digits.

poly p0=3571 - 1580*a^2 - 3075*a*b - 1580*b^2 - 1580*c^2 + 625*a^2*c^2 +
1243*a*b*c^2 + 620*b^2*c^2 - 3075*c*d + 1243*a^2*c*d + 2506*a*b*c*d +
1243*b^2*c*d - 1580*d^2 + 620*a^2*d^2 + 1243*a*b*d^2 + 625*b^2*d^2 +
1000*ev - 1000*a^2*ev - 1986*a*b*ev - 1000*b^2*ev + 1000*ew - 1000*c^2*ew -
1986*c*d*ew - 1000*d^2*ew - 5102*r + 332*a^2*r + 284*a*b*r + 332*b^2*r +
332*c^2*r + 43*a*b*c^2*r + 20*b^2*c^2*r + 284*c*d*r + 43*a^2*c*d*r +
80*a*b*c*d*r + 43*b^2*c*d*r + 332*d^2*r + 20*a^2*d^2*r + 43*a*b*d^2*r -
63*a*b*ev*r - 63*c*d*ew*r + 3644*r^2 + 75*a^2*r^2 + 724*a*b*r^2 + 75*b^2*r^2
+ 75*c^2*r^2 - 401*a*b*c^2*r^2 - 124*b^2*c^2*r^2 + 724*c*d*r^2 -
401*a^2*c*d*r^2 - 1372*a*b*c*d*r^2 - 401*b^2*c*d*r^2 + 75*d^2*r^2 -
124*a^2*d^2*r^2 - 401*a*b*d^2*r^2 + 458*a*b*ev*r^2 + 458*c*d*ew*r^2 -
1301*r^3 - 69*a^2*r^3 - 303*a*b*r^3 - 69*b^2*r^3 - 69*c^2*r^3 +
146*a*b*c^2*r^3 + 42*b^2*c^2*r^3 - 303*c*d*r^3 + 146*a^2*c*d*r^3 +
618*a*b*c*d*r^3 + 146*b^2*c*d*r^3 - 69*d^2*r^3 + 42*a^2*d^2*r^3 +
146*a*b*d^2*r^3 - 139*a*b*ev*r^3 - 139*c*d*ew*r^3 + 185*r^4 + 12*a^2*r^4 +
39*a*b*r^4 + 12*b^2*r^4 + 12*c^2*r^4 - 17*a*b*c^2*r^4 - 4*b^2*c^2*r^4 +
39*c*d*r^4 - 17*a^2*c*d*r^4 - 86*a*b*c*d*r^4 - 17*b^2*c*d*r^4 + 12*d^2*r^4 -
4*a^2*d^2*r^4 - 17*a*b*d^2*r^4 + 13*a*b*ev*r^4 + 13*c*d*ew*r^4;

poly p1=diff(p0,a);
poly p2=diff(p0,b);
poly p3=diff(p0,c);
poly p4=diff(p0,d);
poly p5=diff(p0,ev);
poly p6=diff(p0,ew);
poly p7=diff(p0,r);

//Solution for r=7/5. Definition of the set of polynomial equations as an ideal.
ideal i1=p1,p2,p3,p4,p5,p6,5*r-7;
//For structural optimization, the ideal is given as;
```

```
//ideal i1=p1,p2,p3p4,p5,p6,p7;

//Change in variables by substitution; a=t+s,b=t-s,c=u+v,d=u-v.
ideal i2=subst(i1,a,t+s);
i2=subst(i2,b,t-s);
i2=subst(i2,c,u+v);
i2=subst(i2,d,u-v);
//Print out
i2;

//The change in the ring, in accordance with the change in variables.
//For the purpose of speed up, the degree reverse lexicographical ordering is used,
//specified by "dp" at the end of the definition.
ring R2=0,(s,t,u,v,ev,ew,r),dp;
ideal i2=imap(R,i2);
i2;
option(redSB); //Option for the reduced Grobner bases.
//Grobner bases generation.
ideal J=std(i2);
"Standard basis w.r.t. degree reverse lexicographical ordering";
J;
print( vdim(J) );
print( dim(J) );

//Change in monomial ordering by FGLM.
//The triangulation of Grobner basis is executed with lexicographical
//monomial ordering. The monomial ordering is changed again.
ring R3=0,(s,t,u,v,ev,ew,r),lp;
ideal J=fglm(R2,J);
"Standard basis w.r.t. lexicographical ordering computed by fglm";
J;
print( vdim(J) );
print( dim(J) );
"Decomposition of the zero-dimensional ideal into triangular sets";
list J2= triangM(J);
J2;
print( size(J2));

//Numerical solving.
for(int i=1;i<=size(J2);i++)
{
//Only solves the equation of r in each polynomial triangle;
```

```
"Solution of unknown r in the triangular set No. "+string(i);
poly getr=J2[i][1];
laguerre_solve(getr,50);
}
//Whole solutions of polynomial triangles.
def AC=triang_solve(J2,50,0,"nodisplay");
size(AC);
setring AC;
//A list "rlist" stores solutions.
for (int i=1;i<=size(rlist);i++)
{
"Solution No."+string(i)+"(s,t,u,v,ev,ew,r)";
rlist[i];
}
quit;
```